\documentclass[sigconf, screen, nonacm]{acmart}

\keywords{GPU kernel, power wall, Tensor-core, power-aware kernel optimization, DVFS}

\usepackage{microtype}
\usepackage{graphicx}
\usepackage{subcaption}
\usepackage{enumitem}
\usepackage{booktabs}
\usepackage{array}
\newcolumntype{C}[1]{>{\centering\arraybackslash}p{#1}}
\newcolumntype{L}[1]{>{\raggedright\arraybackslash}p{#1}}
\usepackage{cleveref}

\newcommand{\oursys}{\textsc{Joule}}

\makeatletter
\renewcommand{\@mkauthors}{%
  \global\setbox\mktitle@bx=\vbox{%
    \unvbox\mktitle@bx
    \medskip
    \centering
    \def\and{,\quad}%
    {\@authorfont\authors\par}%
    \smallskip
    {\@affiliationfont
      \textsuperscript{*}Equal contribution\quad
      \textsuperscript{1}Imperial College London\par}%
    {\small\texttt{\{tony.liu20, zhiwen.mo25, hongxiang.fan, a.zhao\}@imperial.ac.uk}\par}%
    \bigskip}}
\makeatother

\begin{document}

\title{Joule is FLOP: Optimizing Tensor-Core Kernels Beyond Latency Hiding at the GPU Power Wall}

\author{\texorpdfstring{Taowen Liu\textsuperscript{*1}}{Taowen Liu}}
\author{\texorpdfstring{Zhiwen Mo\textsuperscript{*1}}{Zhiwen Mo}}
\author{\texorpdfstring{Hongxiang Fan\textsuperscript{1}}{Hongxiang Fan}}
\author{\texorpdfstring{Aaron Zhao\textsuperscript{1}}{Aaron Zhao}}
\affiliation{%
  \institution{Imperial College London}
  \city{London}
  \country{United Kingdom}}
\renewcommand{\shortauthors}{Liu et al.}

\begin{abstract}
For a tensor-core kernel whose average board power stays near the GPU's power cap, runtime is approximately the energy consumed per call divided by that cap.
Data movement whose latency is hidden by computation still consumes energy; reducing this traffic can raise the sustained clock and shorten execution even without reducing cycle count.
We introduce \textbf{\oursys{}}, a framework that accounts for this energy--frequency coupling in tensor-kernel design and scheduling.
Its cost model combines hardware event counts, physical data-path energy costs, and nonlinear voltage scaling across domains to solve for sustained frequency and runtime under a power cap.
Combining the model with data-movement lower bounds across memory levels and dies yields a power-aware speed of light (PSoL) that quantifies structural and implementation gaps.
This analysis guides wide-tile kernels that use the full tensor memory (TMEM) capacity on B200, with fine-grained asynchronous rotation to overlap accumulator readout with tensor-core computation.
It also guides power-aware spatial and temporal tile scheduling that accounts for trade-offs among read sharing, locality, and load balance.
Given measured event counts, the model achieves 0.69\% mean energy prediction error on B200.
Across 60 dense-layer shapes, our GEMM delivers geometric-mean speedups of 4.94\%, 6.64\%, and 2.12\% over exhaustively selected cuBLAS algorithms on B200, H100 NVL, and RTX PRO 6000, respectively.
For fused Mega-MoE on eight B200 GPUs, the resulting optimizations improve performance by 2.60\%--5.17\% over the upstream DeepGEMM baseline with die-local expert weights.
We will open source \oursys{} upon publication.
\end{abstract}

\maketitle

\section{Introduction}

\label{sec:intro}

\begin{figure}[t]
  \centering
  \includegraphics[width=\columnwidth]{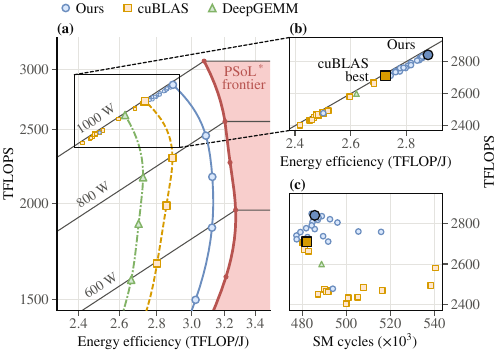}
  \caption{
  $8192^3$ MXFP8 GEMM on B200. (a) Power Roofline with our $\text{PSoL}^*$ frontier (\cref{sec:psol}). 
  We observe that throughput correlates with energy efficiency in (b), while SM cycle count alone in (c) does not reflect this correlation.   \vspace{-1em}
  }
  \label{fig:hero}\label{fig:power_vs_cycles}\label{fig:roofline}\label{fig:power_roofline}

\end{figure}

Tensor-core kernels dominate LLM training and prefill workloads~\citep{narayanan2021megatron,gale2023megablocks,dao2022flashattention}. Kernel optimization overlaps memory access with computation to keep tensor cores busy, reaching roughly 95\% utilization on large GEMMs. These kernels also routinely hit the GPU's board power limit, or \emph{power wall}~\citep{patel2024polca,patel2024splitwise,zhao2023sustainable}. Dynamic voltage and frequency scaling (DVFS) then lowers the clock to stay within the power cap $P_{\text{cap}}$~\citep{guerreiro2019decoupling}. With the cap already at the GPU's maximum, further speedups require kernel designs that use this power budget more efficiently.

\begin{figure}[t]
  \centering
  \includegraphics[width=\columnwidth]{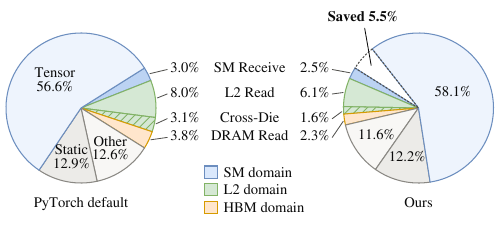}
    \caption{Energy breakdown of one $8192^3$ MXFP8 GEMM on B200 under the 1000\,W cap. 
    Our kernel reduces total energy by 5.5\% over PyTorch default as the data-movement drops from 17.9\% to 12.6\%. 
    \vspace{-1em}
    }
  \label{fig:energy_pie}
\end{figure}

Under the power wall, runtime follows $t \approx E/P_{\text{cap}}$, where $E$ is kernel's energy. Data movement still consumes energy even when computation fully hides its latency: in PyTorch's default cuBLAS kernel on B200, latency-hidden data movement accounts for 17.9\% of total energy (\cref{fig:energy_pie}). Reducing this traffic lowers energy and raises the sustained clock~\citep{kuruppu2026cluster}, so throughput under a fixed cap tracks energy efficiency (\cref{fig:hero}b) while correlating poorly with SM cycle count (\cref{fig:hero}c). On an $8192^3$ MXFP8~\citep{rouhani2023microscaling} GEMM on B200, our kernel executes 0.6\% more cycles than cuBLAS best~\citep{nvidia2026cublas} (\cref{fig:hero}c), yet cuts data-movement energy (\cref{fig:energy_pie}) to consume 5.6\% less energy per call and run 5.19\% faster (\cref{fig:hero}b).

This observation exposes a \emph{power-wall-aware optimization space beyond latency hiding}. Tile size and multicast affect data-movement energy even when computation fully hides data-access latency. Scheduling and implementation also determine energy and cycle losses from redundant traffic, synchronization, and load imbalance. Under $P_{\text{cap}}$, these costs affect the sustained clock, so configurations with similar cycle counts can have different runtimes (\cref{fig:hero}b). Existing kernels on new generation of hardware (e.g. Blackwell GPUs) cannot fully utilize on-chip hardware. Existing auto-tuners~\citep{ansel2014opentuner,vanwerkhoven2019kerneltuner,zheng2020ansor,tillet2019triton} search the configurations exposed by a kernel implementation. Even exhaustive search remains bounded by the schedules that implementation can express.

Exploiting this space requires predicting how a mechanism changes both energy and frequency. Performance models estimate cycles~\citep{huang2014gpumech,lee2025neusight}, while GPU power models estimate power at a given clock~\citep{leng2013gpuwattch,kandiah2021accelwattch,tran2026wattchmen}. A power-capped kernel's clock instead depends on its data paths and their nonlinear voltage-dependent energy costs. Developers also need a speed limit to distinguish losses imposed by the kernel structure from avoidable scheduling overhead. Classical I/O bounds~\citep{hong1981io,irony2004communication,smith2017tight} quantify minimum traffic, but translating these bounds into power-capped speed requires accounting for the energy cost of each physical path (\cref{sec:gaps}).

The kernel must also make energy-saving schedules feasible. Wider tiles reduce input traffic but leave less accumulator space for buffering, which can expose pipeline stalls. Sharing reads across SMs saves energy but adds synchronization and restricts tile alignment. Unequal die speeds further couple locality with load balance, since early-finishing SMs continue to consume idle power. Choosing a schedule requires evaluating these competing effects together (\cref{sec:motivation}).

We present \textbf{\oursys{}}, a framework for power-wall-aware tensor-kernel design and scheduling. \oursys{} connects energy analysis to an expanded design space through four contributions:
\begin{itemize}[leftmargin=*]
    \item \textbf{Energy--frequency cost model.} We model hardware events and per-byte energy costs along physical data paths, account for nonlinear voltage scaling across domains, and solve for the frequency and runtime allowed by $P_{\text{cap}}$. Calibration combines microbenchmarks with real-kernel measurements. With measured event counts, the model achieves 0.69\% mean energy error on B200 (\cref{sec:cost_model,sec:eval_model}).
    \item \textbf{Power-aware speed limits and diagnosis.} We derive multi-level GEMM I/O lower bounds that account for dual-die traffic and combine them with the cost model to obtain a \emph{power-aware speed of light} (PSoL), forming the $\text{PSoL}^*$ frontier in \cref{fig:hero}a. The resulting bounds separate structural losses from implementation losses to guide kernel redesign (\cref{sec:psol,sec:gemm_diag}).
    \item \textbf{Wide-tile asynchronous kernels.} On Blackwell GPUs, we introduce a new kernel structure that dedicates all of TMEM to a single wide tile. Fine-grained asynchronous rotation overlaps accumulator readout with computation without double buffering, cutting input traffic while sustaining tensor-core utilization across dense GEMM and fused Mega-MoE (\cref{sec:gemm_kernel,sec:moe_design}).
    \item \textbf{Power-aware spatial and temporal tile scheduling.} Our tile scheduler uses integer programming to trade off multi-level energy and cycle costs by jointly setting variable tile widths, die assignment, read sharing, and cache residency order (\cref{sec:gemm_planner,sec:moe_design}).
\end{itemize}

Across 60 dense-layer shapes from five Qwen3 models~\citep{yang2025qwen3}, \oursys{}'s GEMM outperforms cuBLAS~13.1's best algorithm selected by exhaustive per-shape search. Geomean speedups are 4.94\% on B200, 6.64\% on H100 NVL, and 2.12\% on RTX PRO 6000 (\cref{sec:eval_gemm}). On eight B200 GPUs, the same design principles accelerate fused Mega-MoE by 2.60\%--5.17\% over the upstream DeepGEMM kernel with die-local expert weights (\cref{sec:eval_moe}). These gains show that expanding the design space can improve on exhaustive tuning within existing implementations.

We plan to open source \oursys{} to help the research community analyze power-wall bottlenecks, develop energy-aware kernel schedules, and build on this design space.

\section{Background and Motivation}
\label{sec:background-motivation}

\subsection{Background}\label{sec:background-hw}

\subsubsection{Tensor Core Kernel Execution, Scheduling, and Auto-Tuning.}

Tensor-core kernels, such as dense GEMM, MoE grouped GEMM, and attention, account for the vast majority of execution time and energy in LLM training and prefill, making even modest speedups highly valuable.
These kernels compute output matrices in tiles~\citep{goto2008anatomy,vanzee2015blis,tomov2010magma}: input tiles are staged from HBM through L2 into on-chip memory~\citep{bose2026tma}, while MMA instructions accumulate partial sums in warp registers (H100 NVL) or dedicated Tensor Memory (TMEM on B200)~\citep{luo2026dissecting,semianalysis2026blackwell}, typically reserving two accumulator buffers (\emph{ping-pong})~\citep{nvidia2026cutlass,pytorch2024pingpong} to overlap epilogue readout with computation.
To maximize data reuse and hide latency, kernels combine (1) \emph{on-chip sharing}, which deduplicates L2 reads across SMs via the GPC request coalescer~\citep{nvidia2026ncu,pan2026async} or SM cluster multicast, and (2) \emph{tile scheduling}, which orders tile execution (L2 swizzling) to keep shared operands in L2 and pipelines asynchronous memory fetches behind MMA cycles.
Because these tiling, sharing, and scheduling choices interact across matrix shapes, state-of-the-art libraries such as cuBLAS, CUTLASS, ThunderKittens, and DeepGEMM~\citep{nvidia2026cublas,nvidia2026cutlass,spector2025thunderkittens,deepgemm} rely on heuristics or black-box \emph{auto-tuning}~\citep{whaley2001atlas,frigo2005fftw3,nugteren2015cltune,vanwerkhoven2019kerneltuner,chen2018learning,shao2022metaschedule,tillet2019triton}---profiling candidate kernels on hardware to pick the fastest configuration per shape---already keeping active tensor cores busy for $\sim$95\% of cycles on large GEMMs.

\subsubsection{Multi-Die Memory GPU} 
Modern datacenter GPUs increasingly adopt multi-die packaging~\citep{arunkumar2017mcmgpu,fan2026diescaling}. For instance, NVIDIA B200 integrates two compute dies connected via a high-bandwidth cross-die link; each die has its own SMs, a 63\,MiB L2 slice, and local HBM. A byte delivered to an SM can thus come from the local L2 (individually or merged via GPC coalescing / cluster multicast), across the cross-die link from the remote L2, or from HBM~\citep{fan2026diescaling,lam2025b200}. By default, physical pages are interleaved across both dies, and cross-die accesses duplicate cache lines in the local L2~\citep{milic2017numa} (halving effective L2 capacity for shared data), though tensors can also be explicitly placed on a specific die via CUDA locality domains~\citep{nvidia2026locality}. 

\subsubsection{Power-Capped Kernels and DVFS}
Tensor-core kernels routinely hit the board power limit $P_{\text{cap}}$ ($\sim$1000\,W on B200, $\sim$400\,W on H100 NVL). Firmware then uses Dynamic Voltage and Frequency Scaling (DVFS) to throttle clock frequencies $f_d$ and supply voltages $v_d(f_d)$ in regulated domains (e.g., SM and L2) until power equals $P_{\text{cap}}$. Consequently, a kernel's sustainable frequency depends on its energy per cycle, and runtime is governed by total energy: $t \approx E / P_{\text{cap}}$.

\subsection{Motivation: Gaps and Challenges Under the Power Wall}
\label{sec:motivation}

\subsubsection{The Hidden Energy Cost of Latency-Hidden Traffic.}
Conventional kernel optimization reasons primarily in execution cycles: as long as memory accesses are hidden behind tensor-core computation without stalling the pipeline, operations off the critical path are treated as free. Under the power wall, however, latency-hidden operations still cost energy and directly depress clock frequency (\cref{fig:hero,fig:energy_pie}). Even in PyTorch's default cuBLAS kernel on an $8192^3$ GEMM on B200, $17.9\%$ of the energy is consumed by data movement that is already hidden behind computation: $3.0\%$ for receiving bytes at the SMs, $8.0\%$ for L2 reads, $3.1\%$ for cross-die transfers, and $3.8\%$ for DRAM reads. Reclaiming this hidden energy allows DVFS to raise clock frequency and finish earlier, though the resulting speedup is nonlinear: as fewer events allow frequency $f_d$ to rise, supply voltage $v_d(f_d)$ climbs steeply along a nonlinear curve, making remaining events in that domain more expensive and partially offsetting the saved energy.

\subsubsection{Gaps in Existing Tools.}\label{sec:gaps}
Existing performance and power tools cannot guide kernel optimization under a power cap:
\begin{itemize}[leftmargin=*,nosep]
    \item \textbf{Hardware Counters and Profilers:} Tools such as Nsight Compute~\citep{nvidia2026ncu} (\texttt{ncu}) report cycle-level metrics and hardware event counts, but cannot attribute board-level power to individual events or physical data paths. Worse, to collect deterministic counters across replays, \texttt{ncu} locks the GPU clock to a fixed base frequency by default, erasing DVFS throttling and misjudging designs that trade slightly more cycles for lower energy and higher power-capped speed.
    \item \textbf{Performance Analysis Models:} Performance models and simulators~\citep{hong2009analytical,huang2014gpumech,lee2022gcom,lee2025neusight,mo2026tilesight,bakhoda2009gpgpusim,khairy2020accelsim,sun2019mgpusim} estimate execution at a given clock. Classical bounds such as Roofline~\citep{williams2009roofline} use a fixed compute ceiling that may be unreachable under power-capped DVFS. Power Roofline~\citep{kim2026powerroofline} relates performance to energy efficiency, while fine-grained energy attribution remains an open gap. Classical I/O lower bounds~\citep{hong1981io,irony2004communication,smith2017tight,ballard2011minimizing,aldaas2022tight,olivry2020automated} quantify minimum traffic, but converting them to a power-capped speed limit requires per-path energy costs. Developers therefore still need a speed limit that separates losses imposed by tile size or sharing configuration from losses caused by redundant traffic and pipeline stalls.
    \item \textbf{GPU Power Models:} Event-based power models~\citep{leng2013gpuwattch,kandiah2021accelwattch,arunkumar2019gpujoule,tran2026wattchmen} solve the inverse problem: they predict power at a given frequency rather than solving for the steady-state frequency and runtime allowed by $P_{\text{cap}}$. Moreover, they are unaware of new GPU microarchitecture or multi-domain nonlinear voltage scaling, resulting in 7\%--36\% errors which cannot be used to guide kernel design.
 
    \item \textbf{Selection Heuristics and Auto-Tuning:} Library selection and empirical auto-tuning~\citep{nvidia2026cublas,vanwerkhoven2019kerneltuner,schoonhoven2022going} can capture energy-induced speed differences through measurements, but remain limited to configurations exposed by existing implementations. Even exhaustive selection cannot recover gains that require changing the kernel structure.
\end{itemize}

\begin{figure*}[t!]
  \centering
  \includegraphics[width=\textwidth]{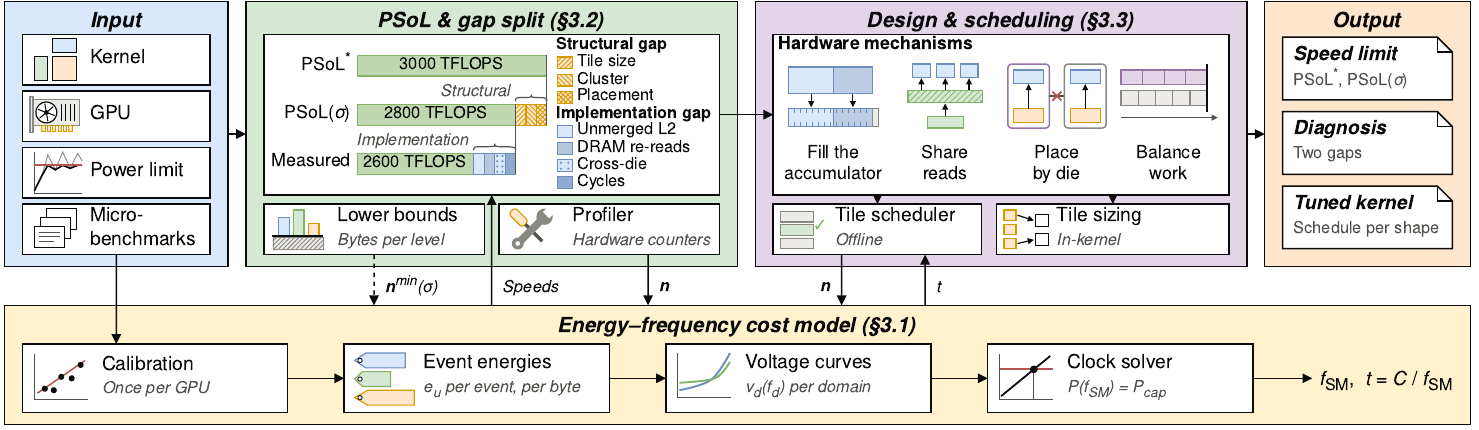}
  \caption{\oursys{} overview. The energy--frequency cost model takes hardware event counts $\vec{n}$ to predict power-capped runtime, derive PSoL speed limits, and guide kernel design and tile scheduling.}
  \label{fig:watt_overview}
\end{figure*}

\subsubsection{Challenges in Power-Wall-Aware Kernel Optimization.}
Even with accurate energy accounting and theoretical bounds, redesigning and scheduling kernels under the power wall faces three tensions:
\begin{itemize}[leftmargin=*,nosep]
   \item \textbf{Trade-off Between Energy and Cycles:} Hardware mechanisms that reduce data movement often introduce extra waiting or synchronization cycles. For instance, using larger multicast clusters reduces L2 read bytes via cross-SM sharing, but forces more SMs to wait for one another at synchronization barriers and worsens tail-wave imbalance.
    \item \textbf{Conflicts Between Optimization Levels:} Energy-saving choices at one memory level frequently constrain optimization at another. For example, maximizing tile width to cut SM received bytes restricts how tiles can be partitioned and aligned for L2 sharing.
    \item \textbf{Cross-Unit Speed Asymmetry:} Conventional scheduling partitions work evenly across compute units, assuming uniform speed. Under the power wall, however, different dies (with asymmetric active SM counts) and different GPUs sustain distinct steady-state frequencies. 
\end{itemize}

\section{\oursys{} Framework}
\label{sec:framework}\label{sec:watt}
\label{sec:workflow}

\Cref{fig:watt_overview} shows the \oursys{} workflow, which connects three components through hardware event counts $\vec{n}$. The \emph{energy--frequency cost model} (\cref{sec:cost_model}) is calibrated once per GPU and converts any event-count vector $\vec{n}$ into steady-state clock frequency and runtime under $P_{\text{cap}}$. \emph{Power-aware Speed of Light (PSoL) and gap decomposition} (\cref{sec:psol}) feed the minimum required event counts $\vec{n}^{\min}(\sigma)$ along each physical path into the cost model to find the speed ceiling under $P_{\text{cap}}$ and compare against measured counts to locate bottlenecks. \emph{Kernel design and tile scheduling} (\cref{sec:guided_design}) modify the kernel to remove the diagnosed bottlenecks and query the cost model to resolve conflicting scheduling choices.

\subsection{Power-Wall-Aware Energy--Frequency Cost Model}
\label{sec:cost_model}

The \oursys{} cost model builds on two core designs. First, it assigns an energy cost to the specific physical path that each data transfer traverses. Second, it scales event energy by the measured nonlinear voltage curve of each clock domain and treats operating frequency as the unknown under $P_{\text{cap}}$ to solve directly for runtime from event counts.

\subsubsection{Energy Modeling and Frequency Solving}
\label{sec:model_formulation}

\paragraph{Multi-Domain Nonlinear Energy Model.}
A GPU contains multiple independent voltage domains, such as the SM, L2, and HBM domains, where units within each domain share a supply voltage and clock frequency. When a kernel hits $P_{\text{cap}}$, DVFS lowers the frequency and voltage of regulated domains, which reduces the per-event energy within those domains. The background energy $\bar{P}t$ comes from constant background power $\bar{P}$ and depends only on execution time $t$. The dynamic energy comes from hardware events, including tensor-core operations, CUDA-core instructions, and bytes transferred along each physical path. For an event $u$ in voltage domain $d$, its per-event energy scales with the squared relative voltage~\citep{chandrakasan1992lowpower} $v_d(f_d)^2$, normalized to $1$ at the maximum frequency:
\begin{equation}
E = \bar{P} t + \sum_{d} v_d(f_d)^2 \sum_{u \in d} e_u n_u,
\label{eq:energy}
\end{equation}
where $f_d$ is the operating frequency of domain $d$, $n_u$ is the count of event $u$, and $e_u$ is its baseline energy cost at the maximum frequency. On modern GPUs, the voltage curve $v_d(f_d)$ is flat at low clocks and steepens sharply near peak frequency. Under $P_{\text{cap}}$, supply voltage drops much less than clock frequency, so assuming constant event energy or linear voltage--frequency scaling causes large prediction errors.

\begin{figure}[t]
  \centering
  \begin{subfigure}[t]{\columnwidth}
    \centering
    \includegraphics[width=\linewidth]{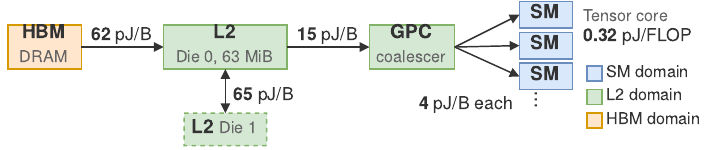}
    \caption{Read paths and per-hop energy costs}
    \label{fig:b200_model_path}
  \end{subfigure}\\[2pt]
  \begin{subfigure}[t]{0.595\columnwidth}
    \centering
    \includegraphics[width=\linewidth]{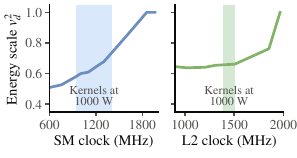}
    \caption{Energy scaling in SM and L2 domains}
    \label{fig:b200_model_voltage}
  \end{subfigure}\hfill
  \begin{subfigure}[t]{0.39\columnwidth}
    \centering
    \includegraphics[width=\linewidth]{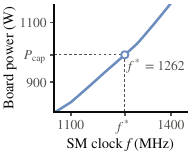}
    \caption{Frequency solving under $P_{\text{cap}}$}
    \label{fig:b200_model_solve}
  \end{subfigure}
  \caption{\oursys{} cost model on B200: (a) physical paths, (b) voltage domain energy scaling $v_d^2$, with the kernel's frequency around 1000\,W shaded, and (c) steady-state clock solving for $8192^3$ GEMM at 1000\,W.}
  \label{fig:b200_model}
\end{figure}

\paragraph{Modeling Architecture-Specific Physical Paths.}
We use microbenchmarks to identify the physical data paths and voltage-domain behaviors that materially affect energy on each GPU architecture. On the dual-die B200 (\cref{fig:b200_model_path}), the model splits L2-to-SM delivery at the GPC request coalescer into an L2 read term and a per-recipient SM receive term to capture the energy saved by shared reads. It also separates local L2 misses by source into cross-die L2 transfers and HBM DRAM reads to capture the impact of dual-die data placement, and scales energy according to our DVFS model under $P_{\text{cap}}$. We apply the same principle to H100 NVL with its split L2 half-slices and to RTX PRO 6000, where the L2 cache and SM share the same voltage domain.

\paragraph{Frequency Solving Under $P_{\text{cap}}$.}
We observe that the L2 clock $f_{\text{L2}}$ and the fixed DRAM clock $f_{\text{DRAM}}$ can both be expressed as functions of the SM clock $f_{\text{SM}}$. Once DVFS links the clock frequencies across domains, the energy in \cref{eq:energy} becomes a single-variable function $E(f_{\text{SM}})$ of the SM clock $f_{\text{SM}}$. For a kernel executing $C$ SM cycles, its runtime is $t(f_{\text{SM}}) = C / f_{\text{SM}}$ and its average power is $P(f_{\text{SM}}) = E(f_{\text{SM}}) / (C / f_{\text{SM}})$. Under $P_{\text{cap}}$, our solver finds the steady-state frequency where average power equals the cap:
\begin{equation}
\frac{E(f_{\text{SM}})}{C / f_{\text{SM}}} = P_{\text{cap}}.
\label{eq:power_cap}
\end{equation}
Because cycles per second and supply voltage both rise with clock frequency, board power increases strictly monotonically with $f_{\text{SM}}$ (\cref{fig:b200_model_solve}). \Cref{eq:power_cap} therefore has a unique root. We solve for $f_{\text{SM}}$ via Brent's method~\citep{brent1973algorithms} to obtain the runtime $C / f_{\text{SM}}$ and energy $E(f_{\text{SM}})$ directly from event counts. This formulation also quantifies the nonlinear offset effect described in \cref{sec:motivation}: reducing hardware events lowers $E(f_{\text{SM}})$ and raises the allowed clock $f_{\text{SM}}$, but the accompanying rise in $v_d(f_d)^2$ partially offsets the saved energy.

\subsubsection{Model Calibration}
\label{sec:model_calibration}

Calibrating event energy costs $e_u$ and voltage curves $v_d(f_d)$ from real kernels alone or from microbenchmarks alone is inaccurate. In real kernels, hardware event counts are strongly collinear. For example, L2 read bytes and SM received bytes scale together across calls, so ordinary least squares cannot separate their individual energy costs~\citep{hoerl1970ridge}. Microbenchmarks isolate one event at a time to break collinearity, but they drive hardware units in isolation at unthrottled clocks, which introduces systematic bias if used directly as model coefficients. To resolve both issues, we fit least squares on real kernels and use Theil mixed estimation~\citep{theil1961pure} to include independent microbenchmark measurements as weighted priors that anchor the collinear terms. Because \cref{eq:energy} is linear in the energy costs $e_u$ and nonlinear in the voltage curves $v_d(f_d)$, we solve this mixed least-squares problem with Variable Projection (VarPro)~\citep{golub1973varpro}. For any given voltage curve parameters, VarPro solves for $e_u$ in closed form via linear least squares~\citep{lawson1995least,stark1995bvls}, leaving only the low-dimensional voltage-curve knots to the outer nonlinear optimizer. Finally, because enabling hardware performance counters changes power draw, we measure steady-state power and frequency with counters disabled~\citep{burtscher2014measuring,yang2024accurate} and collect event counts in a separate profiling pass.

\subsection{Power-Aware Speed of Light (PSoL) and Gap Decomposition}
\label{sec:psol}

Because the cost model takes only event counts as input, it can evaluate existing kernels and also compute the maximum speed any kernel can reach under $P_{\text{cap}}$, along with where a measured kernel loses speed.

\paragraph{Multi-Level Memory and Dual-Die Traffic Lower Bounds.}
For a tensor-core kernel, the algorithm fixes total FLOPs and input sizes. Given hardware capacity limits and a static kernel structure $\sigma$, comprising tile size, multicast cluster width, and die-local operand placement, tile reuse determines the minimum event counts $\vec{n}^{\min}(\sigma)$ required at each memory level:

\textbf{SM received bytes:} Using the largest tile that fits in the on-chip accumulator and buffers maximizes input reuse and minimizes SM received bytes. \textbf{L2 reads:} Assuming full multicast within each cluster and ideal GPC request coalescing across clusters determines the minimum L2 read bytes. \textbf{DRAM reads:} Classical cache-aware I/O lower bounds~\citep{hong1981io, smith2017tight} give the minimum HBM fetches from cold inputs and mandatory re-reads when operands exceed L2 capacity. \textbf{Cross-die reads:} On dual-die GPUs (\cref{fig:b200_model_path}), parallel communication bounds~\citep{irony2004communication} give the minimum cross-die traffic under die-local placement, whereas default page interleaving fixes the cross-die fraction by how work is split between dies.

\paragraph{Structural Gap and Implementation Gap.}
For a static kernel structure $\sigma$, feeding its minimum event counts $\vec{n}^{\min}(\sigma)$ into the cost model (\cref{sec:cost_model}) and solving \cref{eq:power_cap} under $P_{\text{cap}}$ gives the \emph{conditional PSoL}, denoted $\mathrm{PSoL}(\sigma)$. Maximizing $\mathrm{PSoL}(\sigma)$ over all valid structures $\sigma$ gives the GPU's \emph{absolute PSoL}, denoted $\text{PSoL}^*$:
\begin{equation}
\mathrm{PSoL}^* = \max_{\sigma} \mathrm{PSoL}(\sigma).
\label{eq:psol_star}
\end{equation}
Comparing a kernel with measured speed $S_{\text{meas}}$ against both ceilings splits its shortfall from $\text{PSoL}^*$ into two parts:
\begin{equation}
\mathrm{PSoL}^* - S_{\text{meas}} = \underbrace{\big(\mathrm{PSoL}^* - \mathrm{PSoL}(\sigma)\big)}_{\text{Structural Gap}} + \underbrace{\big(\mathrm{PSoL}(\sigma) - S_{\text{meas}}\big)}_{\text{Implementation Gap}}.
\label{eq:gap_decomp}
\end{equation}
\textbf{Structural gap:} $\mathrm{PSoL}^* - \mathrm{PSoL}(\sigma)$. This loss is fixed by the chosen static structure $\sigma$, such as a small tile size, a suboptimal cluster width, or interleaved die placement, and cannot be recovered without changing $\sigma$. \textbf{Implementation gap:} $\mathrm{PSoL}(\sigma) - S_{\text{meas}}$. Once $\sigma$ is fixed, this loss comes from imperfect GPC request coalescing, suboptimal tile traversal that causes extra DRAM or cross-die re-reads, and unhidden pipeline or barrier cycles. Cluster width is fixed at kernel launch and belongs to the structural gap, whereas GPC coalescing depends on which tiles execute concurrently on the same GPC and belongs to the implementation gap. 

\subsection{PSoL Guided Kernel Design}
\label{sec:primitives}\label{sec:guided_design}

Approaching $\text{PSoL}^*$ requires closing both the structural gap and the implementation gap. To close the structural gap, a kernel must use wide tiles to fill the accumulator, share L2 reads across SMs, and place operands in die-local memory. To close the implementation gap, it must also plan the tile schedule, die and GPC assignment, and execution order to cut cache re-reads and keep all SMs finishing together. Because these choices can conflict, such as larger multicast clusters reducing L2 reads but adding barrier waits, \oursys{} uses its cost model to evaluate the trade-offs and pick the schedule with the shortest runtime under $P_{\text{cap}}$. \Cref{sec:design} applies this approach to static-shape GEMM and to dynamically routed Mega-MoE, and \cref{sec:eval_psol} quantifies both gaps.

\section{\oursys{}-Driven Kernel Design and Scheduling}
\label{sec:design}\label{sec:kernels}

Kernels with low tensor-core utilization rarely reach $P_{\text{cap}}$. We therefore apply \oursys{} to two core tensor-core workloads that continuously hit the power wall: static-shape dense matrix multiplication with MXFP8 and FP8 GEMM (\cref{sec:gemm_design}), and dynamically routed fused Mixture-of-Experts with Mega-MoE (\cref{sec:moe_design}).

\subsection{Dense GEMM}
\label{sec:gemm_design}
\begin{figure*}[t]
  \centering
  \includegraphics[width=\textwidth]{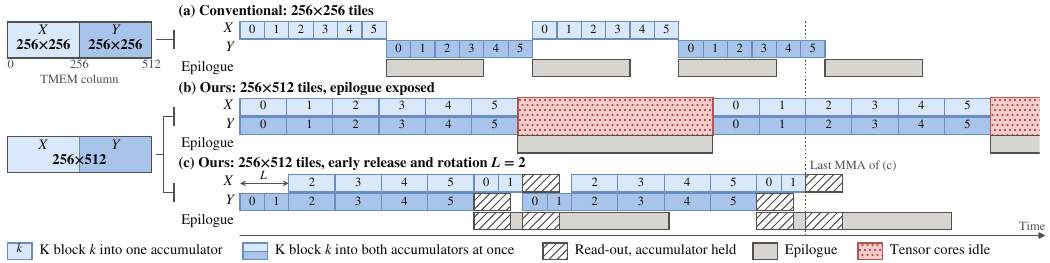}
  \caption{Full-TMEM wide-tile execution on B200, drawn for BF16 with $256\times 512$ tiles across all 512 TMEM columns, whereas MXFP8 reserves 32 columns for scaling factors to form $256\times 480$ tiles. Early TMEM release and $L$-block rotation in (c) eliminate the readout stalls in (b), matching the MMA time of ping-pong buffering in (a) while loading $A$ half as often.
  }
  \label{fig:tile-rotation}
\end{figure*}
We use B200 GEMM as our primary case study. The resulting kernel mechanisms and tile scheduler also apply to GEMM on H100 NVL and RTX PRO 6000 and to fused Mega-MoE on B200 (\cref{sec:moe_b200}).

\subsubsection{Structural Gap Diagnosis and Conditional PSoL}
\label{sec:gemm_bounds}\label{sec:gemm_diag}
We first diagnose the limits of GEMM on B200 (\cref{fig:gemm_diagnosis}, detailed in \cref{sec:eval_psol}). The conditional ceiling $\mathrm{PSoL}(\sigma)$ of a $256\times W$ tile rises with tile width $W$ because wider tiles receive fewer bytes per FLOP at the SMs. With the multicast configurations that cuBLAS best obtains, its $256\times 256$ structure uses only half of TMEM per tile, allowing 94.1\% of $\text{PSoL}^*$ in geomean and reaching 86.3\%, whereas a $256\times 480$ MXFP8 tile allows 96.2\%--97.2\% of $\text{PSoL}^*$. We close this structural gap without losing epilogue overlap via full-TMEM wide tiles with fine-grained asynchronous rotation (\cref{sec:gemm_kernel}), and close the remaining implementation gap from unmerged L2 reads, DRAM and cross-die re-reads, and extra cycles via power-aware tile scheduling (\cref{sec:gemm_planner}).

\subsubsection{Kernel Mechanisms: Full-TMEM Wide Tiles and Fine-Grained Asynchronous Rotation}
\label{sec:gemm_kernel}

\paragraph{Dual-Panel Wide Tiles.}
On B200, a pair of SMs cooperates on a 256-row output band, and \texttt{tcgen05} MMA instructions accumulate products in a 512-column on-chip Tensor Memory (TMEM), shown on the left of \cref{fig:tile-rotation}.\footnote{\Cref{fig:tile-rotation} draws BF16, which requires no scaling factors and uses all 512 TMEM columns to form a $256\times 512$ tile with 256 columns per panel. Our kernel also supports MXFP8, which reserves 32 of the 512 TMEM columns for block scaling factors and caps the maximum tile at $256\times 480$, with up to 256 columns in panel $X$ and 224 in panel $Y$. } After finishing all $K$ blocks of a tile, the epilogue must read the results out of TMEM before the next tile can overwrite those columns. Conventional kernels therefore split TMEM into two 256-column buffers, $X$ and $Y$, and alternate $256\times 256$ tiles between them so one buffer reads out while the other accumulates (\cref{fig:tile-rotation}a). Because only half of TMEM accumulates at any moment, computing the 1024 output columns in \cref{fig:tile-rotation}a requires four tiles and loads operand $A$ four times. Inspired by cooperative scheduling, we instead accumulate into both halves of TMEM at the same time as left and right accumulators $X$ and $Y$ of a single wide tile. In each $K$ step of \cref{fig:tile-rotation}b, both accumulators reuse the same fetched block of $A$ and multiply it by their respective column blocks of $B$. Each wide tile performs twice the work per $K$ block, so two wide tiles finish the 1024 columns in the same total MMA time while loading $A$ only twice. However, using all of TMEM for one tile leaves no spare buffer to hide readout. When both panels finish block~5 together in \cref{fig:tile-rotation}b, the next tile cannot start until the epilogue finishes reading out TMEM, stalling the tensor cores.

\paragraph{Hiding Epilogue via Fine-Grained Asynchronous Rotation.}
We eliminate these stalls using two mechanisms (\cref{fig:tile-rotation}c). First, the epilogue copies a finished panel from TMEM into warp registers and immediately releases its TMEM columns, running the slower HBM writeback asynchronously in the gray region after release. Second, we re-order the $K$ blocks of $X$ from $0, 1, 2, 3, 4, 5$ to $2, 3, 4, 5, 0, 1$, which we call \emph{rotation} ($L=2$ in \cref{fig:tile-rotation}c). $Y$ starts $L$ blocks earlier and executes blocks $0, 1$ alone. Both accumulators then execute blocks $2, 3, 4, 5$ together, sharing a single load of $A$. Finally, $Y$ finishes $L$ blocks earlier and reads out its result while $X$ executes its remaining blocks $0, 1$. This rotation requires loading the first $L$ blocks of $A$ twice, once for each accumulator, while all remaining $K-L$ blocks are loaded once and shared. $L$ is a parameter in the kernel's design space.

\paragraph{Fine-Grained Tile Widths.}
Furthermore, each SM pair reads a static schedule list from memory that specifies the output coordinates and the exact widths of $X$ and $Y$ for each tile. Supporting fine-grained tile widths avoids wasted out-of-bounds computation on indivisible matrix dimensions and exposes per-tile width, cross-die partitioning, SM-to-GPC assignment, and execution order to our tile scheduler (\cref{sec:gemm_planner}).

\begin{figure*}[t]
  \centering
  \begin{subfigure}[t]{0.345\textwidth}
    \centering
    \includegraphics[width=\linewidth]{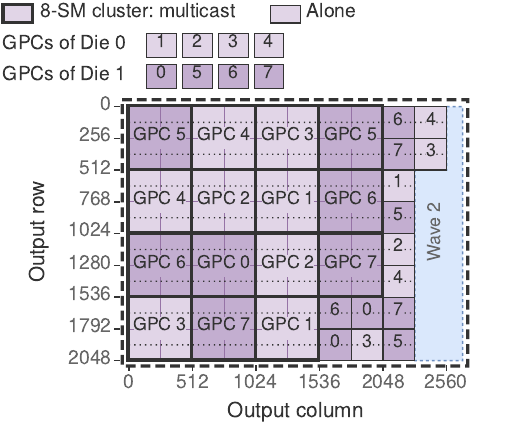}
    \caption{Conventional spatial mapping on Dies 0 \& 1}
    \label{fig:scheduler_conventional}
  \end{subfigure}\hfill
  \begin{subfigure}[t]{0.345\textwidth}
    \centering
    \includegraphics[width=\linewidth]{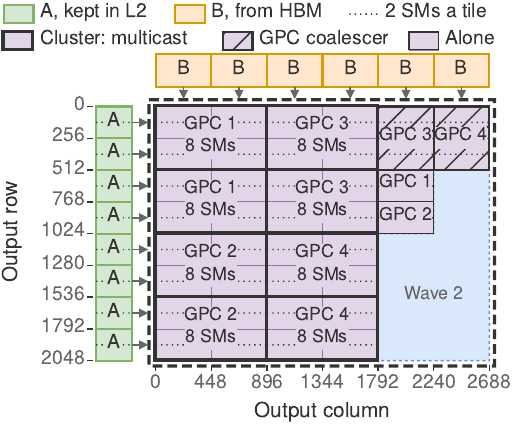}
    \caption{Our spatial mapping on Die 0}
    \label{fig:scheduler_ours}
  \end{subfigure}\hfill
  \begin{subfigure}[t]{0.292\textwidth}
    \centering
    \includegraphics[width=\linewidth]{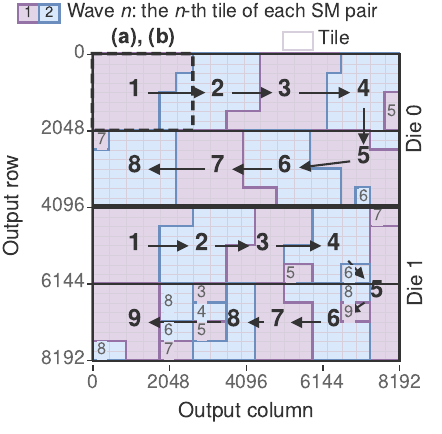}
    \caption{Temporal order across the output and die output split}
    \label{fig:scheduler_time}
  \end{subfigure}
  \caption{Tile schedule for $8192\times 8192\times 16384$ GEMM on B200. Hardware yield floorsweeping leaves GPCs with unequal SM counts that are not always multiples of the CTA cluster size, where GPCs 1,2 have 18 SMs and GPCs 3,4 have 20 SMs. Compared with (a) conventional scheduling, (b) our schedule uses $256\times 448$ wide tiles, keeps 8-SM clusters within one die and GPC, and pairs remainder SMs in the same GPC for request coalescing. (c) Output row split across dies and wave execution order.}
  \label{fig:scheduler}
\end{figure*}

\begin{figure}[t]
  \centering
  \includegraphics[width=\columnwidth]{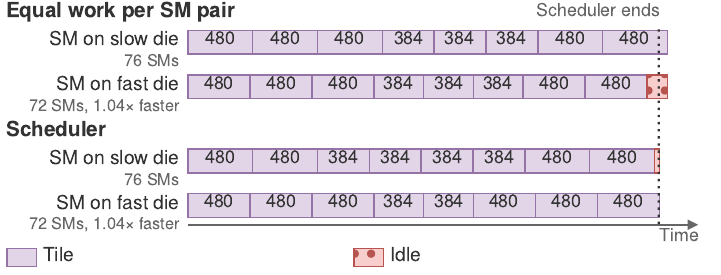}
  \caption{Speed-proportional work assignment across B200's two dies. Each die has a similar power budget, so a die with more SMs has a lower power budget per SM and therefore runs at a lower frequency.}
  \label{fig:scheduler_speed}
\end{figure}

\subsubsection{Power-Aware Tile Scheduler.}
\label{sec:tile_scheduler}\label{sec:gemm_planner}\label{sec:planner}
Our kernel exposes a novel power-aware tile-scheduling design space.
Conventional schedules do not plan tiles according to dies and GPCs (\cref{fig:scheduler}a), which causes cross-die traffic and misses coalescing via the GPC request coalescer for SMs left outside CTA clusters. They also leave the faster die idling at the end of the kernel (\cref{fig:scheduler_speed}, top).
Given a kernel configuration, our scheduler solves for the tile schedule analytically in four steps:
First, to avoid cross-die traffic and match cross-die speed, the scheduler partitions the output workload across the two dies in proportion to measured per-die throughput~\citep{beaumont2002partitioning} (\cref{fig:scheduler}c) and pins each die's input partition in die-local memory~\citep{nvidia2026locality}.
Second, once each die's workload is fixed, an integer program~\citep{huangfu2018highs} decomposes the output dimension into variable tile widths to minimize out-of-bounds wasted work, while assigning wider tiles to SM pairs on the faster die so both dies finish together (\cref{fig:scheduler_speed}, bottom).
Third, to maximize intra-GPC read sharing (\cref{fig:scheduler}b), our scheduler uses both multicast and the GPC request coalescer to merge L2 reads. It spatially packs SMs that remain outside multicast clusters.
Fourth, to maximize L2 reuse (\cref{fig:scheduler}c), the scheduler groups contiguous tiles up to the single-die L2 capacity and traverses them in serpentine wave order so the shared operand stays resident in L2 across waves.

\subsubsection{Kernel Selection.}
Our kernel configurations create a large number of combinations, so we use a heuristic to prune configurations. To evaluate these configurations without GPU runs, a CPU model predicts each candidate's hardware event vector $\vec{n}$ from its synthesized tile schedule, and the \oursys{} cost model (\cref{sec:cost_model}) predicts its power-capped runtime. We can either select the top-ranked candidate directly on the CPU with zero GPU runs, or narrow the candidates to a small top subset for on-device steady-state verification.

\subsubsection{Cross-Architecture Adaptation: H100 NVL and RTX PRO 6000}
\label{sec:gemm_other_gpus}

On H100 NVL and RTX PRO 6000, warp-register accumulators cap tiles at $256\times 128$, so our gains come from adapting the power-aware tile scheduler and cost-model-guided selection (\cref{sec:eval_gemm}). \textbf{H100 NVL:} Our kernel builds with $256\times 128$ tiles and bidirectional multicast. We use our tile scheduler and use the H100 NVL cost model to select the cluster mode, half-slice L2 group size, and traversal order. \textbf{RTX PRO 6000:} On this single-die GPU without cluster multicast or cross-SM coalescing, we employ our tile scheduler and uses the RTX PRO 6000 cost model to trade off $256\times 128$ against narrower tiles for tail-wave balance and to select the L2 swizzle order that minimizes DRAM re-reads.

\subsection{Mega-MoE}
\label{sec:moe_b200}\label{sec:moe_design}
Our power-wall-aware kernel tuning methodology and wide-tile mechanisms extend beyond standalone GEMM to other power-capped tensor-core workloads. We target the expert-parallel MoE layer of DeepSeek-V4-Pro~\citep{deepseek2026v4}. Analyzing the baseline kernel's event counts with \oursys{} shows that its shortfall from $\mathrm{PSoL}^*$ likewise comes from a structural gap due to narrow effective tiles and unshared weight reads, and an implementation gap due to padded remainder blocks and TMEM release stalls. We introduce four runtime improvements while preserving strict batch-invariant outputs:

\textbf{Wide Effective Tile and Rotation.} Following our wide-tile GEMM design (\cref{sec:gemm_kernel}), each SM computes two weight sub-blocks against one fetched token tile. A token tile then serves twice as many weight rows, which halves the token bytes an SM receives per output and lowers $\vec{n}^{\min}(\sigma)$ to close the structural gap. Because two sub-blocks per SM require two accumulators, we rotate the two sub-block accumulators to hide epilogue readout without inflating the implementation gap.

\textbf{Deeper Weight Pipeline.} A step over two sub-blocks consumes two weight stages, so a six-stage weight pipeline no longer hides load latency and widens the implementation gap. We deepen the weight pipeline to eight stages and free the shared memory for them by shortening the token pipeline and pulling half a token tile at a time.

\textbf{In-Kernel Tile Size Selection.} Routing determines each expert's token count only at run time, so the kernel selects tile sizes in-kernel: it splits each expert's tokens evenly into tiles whose heights are multiples of 16, which removes the baseline's padded remainder tile and its wasted MMA cycles. Eliminating these padded tiles cuts both wasted tensor-core energy and extra cycles, directly shrinking the implementation gap under dynamic routing.

\textbf{Weight Sharing.} The two SM pairs of a 4-SM cluster claim adjacent token tiles of the same expert and weight column, so that one weight tile loaded from L2 is multicast to both pairs. This cluster multicast halves the minimum L2 weight bytes in $\vec{n}^{\min}(\sigma)$ and further narrows the structural gap to $\mathrm{PSoL}^*$.

As in the upstream kernel, each die processes half of every weight block from die-local memory. Rotation fixes the accumulation order along $K$ for each projection, so outputs stay batch-invariant; they differ from the upstream kernel's only in that order. We leave the co-design of token routing and tile scheduling to future work.

\section{Evaluation}
\label{sec:eval}

We evaluate \oursys{} around four questions. First, how accurately does the cost model predict power-capped energy, runtime, and relative speedups across unseen kernels in \cref{sec:eval_model}? Second, does our \oursys{}-tuned dense GEMM outperform state-of-the-art libraries across GPU architectures, and how does our kernel selection compare against existing auto-tuners and generic tuners in \cref{sec:eval_gemm}? Third, do the same design principles generalize to dynamically routed, multi-GPU fused Mega-MoE in \cref{sec:eval_moe}? Fourth, how much of the $\text{PSoL}^*$ gap does each mechanism close, and how much does each mechanism contribute in \cref{sec:eval_ablation}?

\subsection{Evaluation Setup}
\label{sec:eval_setup}\label{sec:model-measure}

We evaluate \oursys{} on NVIDIA B200 at 1000\,W, H100 NVL at 400\,W, and RTX PRO 6000 at 600\,W using CUDA~13.4 and cuBLAS~13.1. We test MXFP8 GEMM on B200 and RTX PRO 6000, $1\times 128$ FP8 GEMM~\citep{micikevicius2022fp8} on H100 NVL, and BF16 GEMM on B200 and H100 NVL across all 60 dense-layer shapes from five Qwen3 models~\citep{yang2025qwen3,alibaba2026qwen38max} at $M \in \{2\text{K}, 4\text{K}, 8\text{K}, 16\text{K}\}$, plus DeepSeek-V4-Pro~\citep{deepseek2026v4} fused Mega-MoE (\cref{sec:eval_moe}). We report geomean speedups against \emph{PyTorch}'s default cuBLAS heuristic, \emph{cuBLAS best} via exhaustive per-shape sweep, and \emph{DeepGEMM}~\citep{deepgemm}. Using normally distributed inputs~\citep{gregersen2024input,he2024predictable} and L2 flushes, each point is the median of three 3\,s CUDA Graph runs from a $45\,^\circ\text{C}$ cooldown, discarding the first 1\,s to measure DVFS steady state (\cref{sec:eval_boost} evaluates the initial boost window).

\subsection{Energy--Frequency Cost Model Evaluation}
\label{sec:eval_model}

\begin{figure}[t]
  \centering
  \includegraphics[width=\columnwidth]{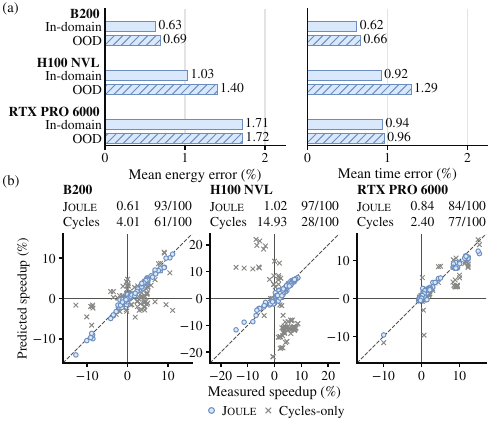}
  \vspace{-1em}
  \caption{\oursys{} cost-model accuracy across three GPUs. Single-kernel in-domain and OOD errors stay near 1\%, and pairwise ranking outperforms cycle ratios.}
  \vspace{-1em}
  \label{fig:model-accuracy}\label{tab:model-accuracy}
\end{figure}

This section evaluates the \oursys{} cost model using measured hardware event counts $\vec{n}$, while \cref{sec:eval_planner} evaluates it jointly with analytical event-count estimation for kernel selection. Each GPU requires only a one-time calibration, and all results are reported on held-out data.

\paragraph{Energy and Runtime Accuracy}
\Cref{fig:model-accuracy}a reports energy and power-capped runtime prediction errors across the three GPUs under two splits. The \emph{in-domain} split performs 10-fold cross-validation across real-kernel measurement points, whereas the \emph{out-of-domain} (OOD) split excludes an entire algorithm family from training, such as DeepGEMM, wide-tile GEMM, fused Mega-MoE, or FlashAttention~\citep{zadouri2026flashattention4}. Across both splits and all three GPUs, mean energy and runtime errors stay around 1\%. Errors on RTX PRO 6000 are slightly higher because the card switches between two voltage curves with temperature, preventing short microbenchmarks from reaching the voltage regime of heavy kernels and leaving static and memory-path terms anchored mainly by priors.

\paragraph{Pairwise Speedup Ranking}
\Cref{fig:model-accuracy}b compares the \oursys{} cost model against cycle ratios alone across 100 held-out candidate pairs per GPU. On B200 and H100 NVL, \oursys{} identifies the faster kernel much better than cycle ratios. On RTX PRO 6000, \oursys{} cuts rms error from 2.40 to 0.84 percentage points and hits 48 of 49 pairs differing by more than 1 percentage point.

\begin{table}[t]
  \centering
  \small
  \setlength{\tabcolsep}{4.5pt}
  \caption{Model comparison and structural ablation on B200. Each row in the bottom block adds one feature to the preceding row. \oursys{}'s energy errors are the held-out-family errors of \cref{fig:model-accuracy}a; the other rows leave one kernel out.}
  \label{fig:model-ablation}\label{tab:model-ablation}
  \begin{tabular}{@{}lrrrr@{}}
    \toprule
     & \multicolumn{2}{c}{Energy error (\%)} & \multicolumn{2}{c}{Pairwise speedup} \\
    \cmidrule(lr){2-3}\cmidrule(l){4-5}
    Model & Mean & Max & rms (pp) & Hits \\
    \midrule
    Fixed energy per FLOP & 5.89 & 25.14 & -- & -- \\
    Fixed energy per cycle & 3.68 & 13.52 & 4.01 & 61/100 \\
    \midrule
    Wattchmen~\citep{tran2026wattchmen} & 21.27 & 30.42 & 2.70 & 74/100 \\
    AccelWattch~\citep{kandiah2021accelwattch} & 7.23 & 23.02 & 2.66 & 78/100 \\
    \midrule
    No-$v(f)^2$ base & 2.32 & 10.49 & 2.25 & 78/100 \\
    \quad + $V$--$f$ curves & 0.95 & 3.97 & 1.97 & 83/100 \\
    \quad + GPC Coalescer & 0.82 & 4.49 & 1.42 & 87/100 \\
    \quad + Locality aware (\oursys{}) & \textbf{0.69} & \textbf{2.39} & \textbf{0.61} & \textbf{93/100} \\
    \bottomrule
  \end{tabular}
\end{table}

\Cref{tab:model-ablation} compares \oursys{} against heuristic baselines and existing power models on B200, and ablates the components from \cref{sec:model_formulation,sec:model_calibration}. Wattchmen~\citep{tran2026wattchmen} calibrates solely on microbenchmarks at boost clock and overestimates power-capped energy by 21.27\%. AccelWattch~\citep{kandiah2021accelwattch}, even when refitted with our calibration procedure, assumes linear voltage--frequency scaling and incurs 7.23\% error on B200 and 36\% on H100 NVL. Refitting Wattchmen's constant-energy formulation with our Theil mixed calibration (\texttt{No-$v(f)^2$ base}) drops error to 2.32\% because power-capped kernels operate within a narrow clock band, whereas omitting microbenchmark priors raises worst-case OOD error from 2.70\% to 3.57\%. Adding measured nonlinear $v_d(f_d)^2$ curves (\texttt{+ $V$--$f$ curves}) provides the largest single-call error drop to 0.95\%, while modeling the GPC request coalescer and dual-die locality (\texttt{+ GPC Coalescer} and \texttt{+ Locality aware}) drives the pairwise ranking gain, reducing rms error to 0.61 percentage points and energy error to 0.69\%.

\subsection{GEMM Evaluation}
\label{sec:eval_gemm}

\begin{figure*}[t!]
  \centering
  \includegraphics[width=\textwidth]{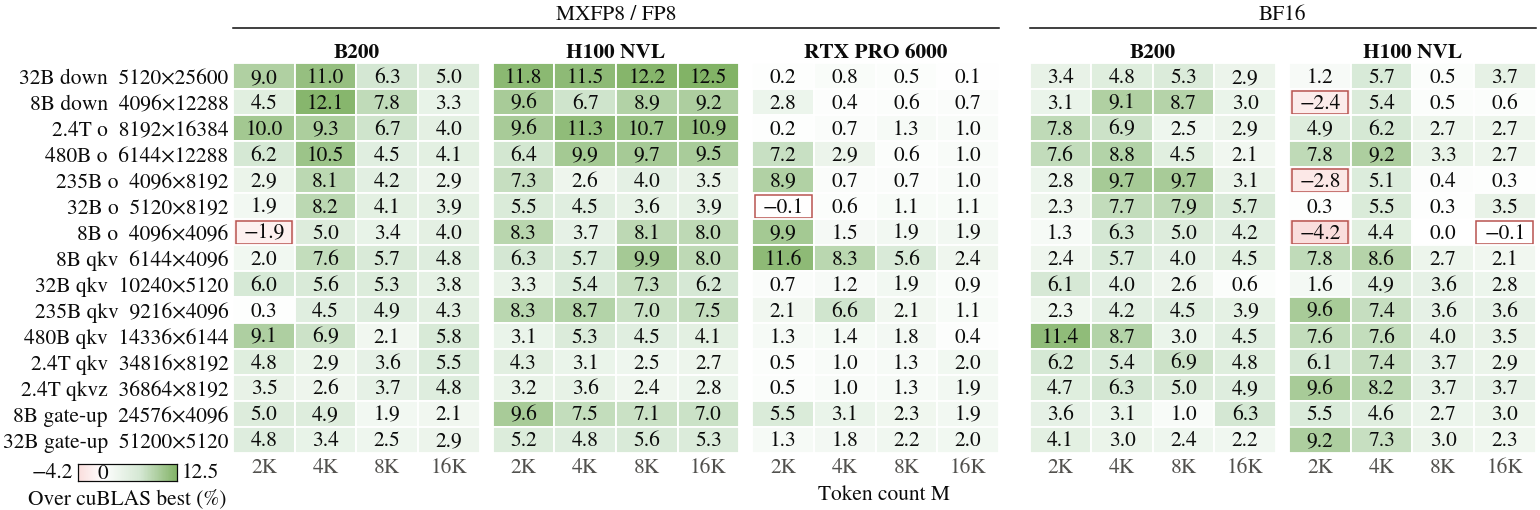}
  \caption{
    Per-shape speedup (\%) of our GEMM over cuBLAS best. We outperform the SOTA on 294 out of 300 shapes.
  }
  
  \label{fig:main-speedups}\label{fig:bf16-speedups}
\end{figure*}

\subsubsection{GEMM Results}
\label{sec:eval_gemm_main}

\begin{figure}[t]
  \centering
  \begin{subfigure}[t]{\columnwidth}
    \centering
    \includegraphics[width=\linewidth]{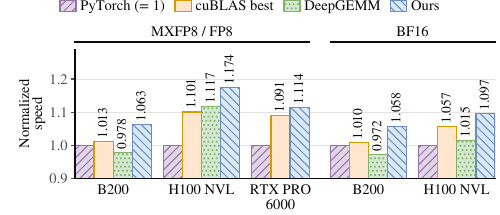}
    \caption{Over the shapes of \cref{fig:main-speedups}}
    \label{fig:main-geomean}
  \end{subfigure}
  \begin{subfigure}[t]{\columnwidth}
    \centering
    \includegraphics[width=\linewidth]{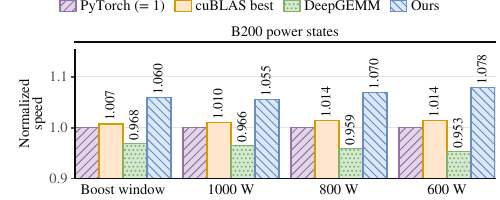}
    \caption{On B200 under different power caps}
    \label{fig:boost}
  \end{subfigure}
  \caption{Geomean speed normalized to PyTorch. In (b), both cuBLAS best and our schedule are selected for each power cap.}
  \vspace{-1em}
  \label{fig:geomean-bars}
\end{figure}

\paragraph{Main Results}
\label{sec:eval-gpus}
\Cref{fig:main-speedups,fig:main-geomean} show that our GEMM consistently outperforms all baselines across the three GPUs. Over cuBLAS best, PyTorch, and DeepGEMM, our kernel achieves geomean speedups of 4.94\% with wins on 59 of 60 shapes, 6.26\%, and 8.64\% on \textbf{B200}. On \textbf{H100 NVL}, it achieves speedups of 6.64\% winning all 60 shapes, 17.40\%, and 5.12\%. On \textbf{RTX PRO 6000}, it achieves speedups of 2.12\% winning 59 of 60 shapes and 11.44\% over cuBLAS best and PyTorch. Gains peak on large-$K$ shapes, whereas the few regressions occur only at $M=2048$, where each SM pair processes roughly one tile and cannot amortize launch overhead.

\noindent\paragraph{Impact of Die-Local Memory Placement}
\label{sec:eval_placement}
Because cuBLAS does not place operands into die-local physical memory as described in \cref{sec:gemm_planner}, we evaluate our B200 GEMM both with and without caller-side die placement. Even under default page-interleaved allocation without placement, our kernel beats cuBLAS best on 58 of 60 shapes by 4.25\% geomean. Enabling die-local placement eliminates cross-die traffic for the partitioned operand and raises the geomean speedup to 4.94\%, realizing $\sim$49\% of the theoretical dual-die bound gain.

\paragraph{Boost and Power capped cases}
\label{sec:eval_boost}
During the first few hundred milliseconds when transitioning from idle to full load, B200 allows instantaneous power to rise to roughly 1100\,W, making each kernel 13\%--21\% faster than in steady state. Measurements show that this boost window runs at roughly 1393\,MHz. Kernels operate under a higher effective power cap where runtime is still governed by energy. Across all the power caps in \cref{fig:boost}, we consistently beat all the baselines in geomean over 30 shapes: by 4.4\%--6.4\% over cuBLAS best, 5.5\%--7.8\% over PyTorch, and 9.3\%--13.2\% over DeepGEMM.

\subsubsection{Design Space, Scheduler, and Kernel Selection Evaluation}
\label{sec:eval_planner}

\begin{figure}[t]
  \centering
  \begin{subfigure}[t]{\columnwidth}
    \centering
    \includegraphics[width=\linewidth]{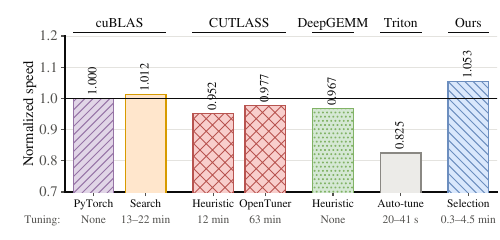}
    \caption{Each tuner in its own design space}
    \label{fig:tuner-eval}\label{tab:tuner-compare}\label{fig:planner-eval}
  \end{subfigure}
  \begin{subfigure}[t]{\columnwidth}
    \centering
    \includegraphics[width=\linewidth]{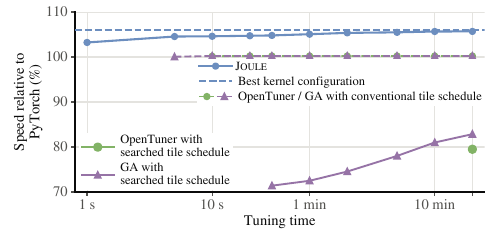}
    \caption{Tuning trajectories in our design space}
    \label{fig:tuner-budget}\label{fig:planner-panels}
  \end{subfigure}
  \caption{Geomean MXFP8 GEMM speed on B200 over 15 $M = 8192$ shapes, normalized to PyTorch.}
  \label{fig:tuner}
\end{figure}

\paragraph{Design Space Evaluation.}
\Cref{fig:tuner-eval} shows that a tuner's performance is bounded by the design space its kernel exposes. Existing libraries and compilers search only the kernel configuration. Because these fixed mappings leave cross-die traffic, L2 re-reads, and tail-wave imbalance unresolved, even 13--63\,min of exhaustive on-device search in cuBLAS or CUTLASS plateaus at $+1.2\%$ over PyTorch default. Exposing per-tile widths, SM-pair mapping, and execution order in our design space lifts the ceiling to $+6.1\%$ and allows our kernel selection to reach $+5.0\%$ using CPU prediction alone.

\paragraph{Scheduler and Kernel Selection vs.\ Generic Tuners.}
\Cref{fig:tuner-budget} shows that expanding the design space alone is insufficient without analytical scheduling. When restricted to the conventional tile schedule, generic tuners such as OpenTuner and a genetic algorithm (GA) search only the kernel configuration and plateau at $+0.2\%$. Conversely, when searching the raw tile schedule directly, both generic tuners face thousands of coupled per-tile variables where uncoordinated assignments destroy L2 locality and cross-die balance, leaving them at $-20.5\%$ and $-17.1\%$ after 20\,min of GPU profiling. Moreover, short on-device timing runs misrank steady-state power-capped performance, whereas full steady-state profiling requires seconds per candidate. 

\paragraph{Scheduling Overhead and Selection Quality.}
For one configuration, analytical scheduling finishes within 0.03--0.42\,s on a single CPU thread. We reach $97.3\%$ within one second, and we are $99.3\%$ of the best selection in our kernel configuration space within 2 minutes.

\subsection{Mega-MoE Evaluation}
\label{sec:eval_moe}\label{sec:multicard}

\begin{figure}[t]
  \centering
  \includegraphics[width=\columnwidth]{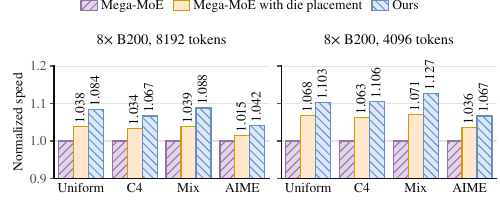}
  \caption{Mega-MoE speed on DeepSeek-V4-Pro over 8 B200 cards.}
  \label{fig:moe}\label{tab:moe}
\end{figure}

We evaluate our fused Mega-MoE kernel from \cref{sec:moe_design} on DeepSeek-V4-Pro under expert parallelism across four routing distributions in \cref{tab:moe}: real traces from a mixed corpus, C4~\citep{raffel2020t5}, and AIME prompts, plus balanced uniform routing.
Across eight B200 GPUs our kernel outperforms the upstream DeepGEMM baseline with die-local expert weights by up to 5.17\%. The speedups come almost entirely from lowering energy to sustain a higher power-capped clock. At 8192 tokens per rank under real routing, the cycle count changes only $-0.5\%$ to $+0.2\%$, whereas its SM frequency rises by 2.5\%--4.6\%.

\subsection{$\text{PSoL}^*$ Gap Analysis and Ablation Study}
\label{sec:eval_ablation}

\subsubsection{$\text{PSoL}^*$ Gap Analysis}
\label{sec:eval_psol}\label{sec:eval-sol}

\begin{figure}[t]
  \centering
  \includegraphics[width=\columnwidth]{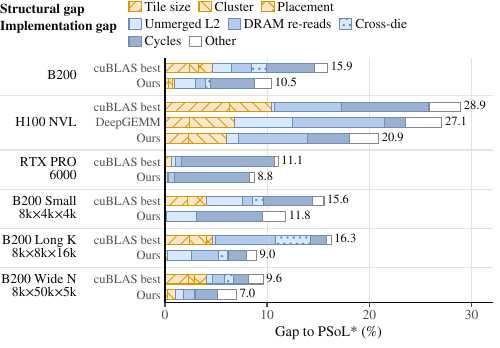}
  \caption{Gap to $\text{PSoL}^*$ across the 60-shape geomean on each GPU and three representative B200 shapes. Our GEMM shrinks both structural and implementation losses across architectures and bottleneck regimes.}
  \label{fig:gemm_diagnosis}
\end{figure}

\Cref{fig:gemm_diagnosis} decomposes each kernel's runtime gap to $\text{PSoL}^*$ into structural and implementation losses, and \cref{tab:traffic} summarizes the underlying changes in cycles, traffic, and energy.

\begin{table}[t]
  \centering
  \footnotesize
  \caption{Geomean change (\%) of our GEMM against each baseline across 60 shapes in SM cycles, transferred bytes per level, and energy.}
  \label{tab:traffic}
  \setlength{\tabcolsep}{2pt}
  \begin{tabular}{@{}lrrrrrrr@{}}
    \toprule
    Ours against & Cycles & Bytes & Receive & L2 & X-die & DRAM & Energy \\
    \midrule
    B200, cuBLAS best & $-1.1$ & $-19.4$ & $-18.7$ & $-16.0$ & $-63.2$ & $-22.7$ & $-4.9$ \\
    H100 NVL, cuBLAS best & $-10.1$ & $-20.4$ & $-24.6$ & $-9.2$ & -- & $-1.2$ & $-6.1$ \\
    H100 NVL, DeepGEMM & $+0.7$ & $-15.0$ & $-0.5$ & $-39.8$ & -- & $-11.8$ & $-5.1$ \\
    RTX PRO 6000, cuBLAS best & $-1.7$ & $-3.5$ & $-3.5$ & $-3.6$ & -- & $+3.1$ & $-2.1$ \\
    \bottomrule
    \end{tabular}
  \vspace{-0em}

\end{table}

On \textbf{B200 GEMM}, wide tiles and die-local placement cut the structural gap from 4.7\% for cuBLAS best to 0.9\%, and the tile scheduler lowers the implementation gap from 11.3\% to 9.5\%, which raises $\text{PSoL}^*$ attainment from 86.3\% to 90.5\%. As \cref{tab:traffic} shows, our GEMM cuts SM cycles by only 1.1\%, yet reduces total data movement by 19.4\% and energy by 4.9\%. Across the three representative shapes, the dominant loss shifts from unmerged L2 reads and extra cycles on the small shape to DRAM and cross-die re-reads on the long-$K$ shape, and our design addresses both regimes.

On \textbf{H100 NVL GEMM}, both our kernel and DeepGEMM use $256\times 128$ tiles to shrink the structural gap relative to cuBLAS best's $128\times 128$ tiles, while our scheduler cuts L2 reads by 39.8\% and DRAM reads by 11.8\% over DeepGEMM at nearly identical cycle counts. The remaining implementation gap stems from large-$K$ operands exceeding one L2 half-slice and the lack of simultaneous GPC coalescing for both operands.

On \textbf{RTX PRO 6000 GEMM}, which lacks TMEM and cross-SM sharing, both kernels use register-limited $256\times 128$ tiles with under 1\% structural gap, so cycle overhead dominates the shortfall. Selecting narrower tiles on indivisible shapes and adjusting L2 traversal order when working sets exceed L2 capacity raises $\text{PSoL}^*$ attainment from 90.0\% to 91.9\%.

On \textbf{B200 Mega-MoE}, the eight-stage weight pipeline and 16-aligned in-kernel tile sizing shrink the implementation gap by cutting inter-step idle time from 19\% to 12\% at 4096 tokens per rank. On a single B200 with uniform routing, our kernel raises $\text{PSoL}^*$ attainment from 81.3\%  to 86.1\% at 8192 tokens per rank, and from 80.9\% to 84.7\% at 4096.

\subsubsection{Ablation Study}
To isolate each mechanism, \cref{tab:ablation} ablates our full design on B200 $8192^3$ mxfp8 GEMM. Every component contributes to the 5.19\% speedup, with cross-die output splitting, die-local placement of $A$, and IP load balancing yielding the largest gains. Across all ablated variants except uniform tile assignment, cycle counts stay within 3.7\%, and the variant with the fewest cycles at $0.475\,\mathrm{M}$ is the slowest under $P_{\text{cap}}$ because it consumes the most energy. Individual losses exceed 5.19\% when summed because wide tiles depend on rotation to hide epilogue stalls and IP balancing to avoid tail-wave idle power.
\begin{table}
  \centering
  \small
  \caption{Ablation study of \oursys{}'s mechanisms compared with cuBLAS best.}
  \label{tab:ablation}
  \setlength{\tabcolsep}{4.5pt}
  \begin{tabular}{@{}lrrr@{}}
    \toprule
    Configuration & vs.\ cuBLAS best & Energy (mJ) & Cycles (M) \\
    \midrule
    Full design & $+5.19\%$ & 381.0 & 0.485 \\
    $-$ rotation & $+4.20\%$ & 385.3 & 0.493 \\
    $-$ GPC grouping & $+4.38\%$ & 384.3 & 0.484 \\
    $-$ L2 residency for $A$ & $+3.68\%$ & 386.2 & 0.483 \\
    $-$ die-local placement of $A$ & $+2.26\%$ & 392.9 & 0.483 \\
    $-$ Uniform tile assignment & $+1.75\%$ & 391.7 & 0.514 \\
    $-$ cross-die output split & $+1.48\%$ & 396.2 & 0.475 \\
    \midrule
    cuBLAS best & -- & 403.4 & 0.482 \\
    \bottomrule
  \end{tabular}
\end{table}

\section{Limitations}
\label{sec:limits}

\noindent\emph{Power-Capped Scope.}
\oursys{} accelerates kernels only when they run at $P_{\text{cap}}$ and data movement consumes a substantial share of energy. Kernels below $P_{\text{cap}}$ gain no clock speedup from cutting latency-hidden traffic, while kernels already near the data-movement lower bound gain little from \oursys{}. \emph{Per-GPU Calibration.}
Calibrating the energy--frequency model requires a one-time microbenchmark suite and profiled kernel runs for each GPU architecture to measure voltage curves and fit physical-path energy costs. Applying the model to another card of the same GPU model still requires fitting per-device coefficients.
\emph{Input Data Distribution.}
The model must be calibrated under the target input data distribution because kernel power depends on operand values~\citep{gregersen2024input,he2024predictable}.

\section{Related Work}
\label{sec:related}

\paragraph{GPU Performance and Power Modeling.}
Analytical performance models~\citep{hong2009analytical,huang2014gpumech,lee2022gcom,lee2025neusight,mo2026tilesight} and cycle-level simulators~\citep{bakhoda2009gpgpusim,khairy2020accelsim,sun2019mgpusim} predict execution cycles at a fixed clock, ignoring that the operating frequency under $P_{\text{cap}}$ depends on kernel energy. Event-based GPU power models~\citep{leng2013gpuwattch,kandiah2021accelwattch,arunkumar2019gpujoule,tran2026wattchmen,hong2010ipp,su2014ppep,lee2026energaizer}, DVFS-aware models~\citep{guerreiro2019decoupling}, and energy rooflines~\citep{choi2013roofline,ilic2017beyond} estimate power at a given frequency. However, they overlook modern physical paths such as GPC request coalescing and dual-die interconnects, neglect multi-domain nonlinear voltage scaling, and do not solve for the steady-state clock allowed by $P_{\text{cap}}$. System-level power-capping frameworks~\citep{you2023zeus,choi2023envpipe,chung2024perseus,patel2024polca,patel2024splitwise,stojkovic2025dynamollm} adjust batch sizes, power limits, or job placement, while Kuruppu et al.~\citep{kuruppu2026cluster} observe that reducing traffic raises clocks on H100 NVL, yet all of them treat the kernel implementation as fixed. In contrast, \oursys{} models per-byte energy cost across physical paths and nonlinear domain voltage curves to solve directly for $P_{\text{cap}}$-constrained runtime, achieving almost 1\% energy error and establishing theoretical speed ceilings ($\text{PSoL}^*$ and $\mathrm{PSoL}(\sigma)$) to guide kernel design (\cref{sec:eval_model}).

\paragraph{GPU Tensor-Core Libraries and Multi-Die Locality.}
Modern tensor-core libraries and DSLs such as cuBLAS, CUTLASS, QuACK, ThunderKittens, and DeepGEMM~\citep{nvidia2026cublas,nvidia2026cutlass,daoailab2026quack,spector2025thunderkittens,deepgemm}, along with FlashAttention~\citep{dao2022flashattention,shah2024flashattention3,zadouri2026flashattention4} and persistent megakernels~\citep{chang2024flux,spector2025megakernel,cheng2025mpk}, optimize tile sizes, cluster multicast, and software pipelines to hide memory latency behind MMA cycles. Because they treat latency-hidden traffic as free, these designs leave half of B200's TMEM idle in $256\times 256$ ping-pong buffers and suffer from redundant L2 and DRAM reads under $P_{\text{cap}}$. Multi-chiplet and NUMA-aware GPU studies~\citep{arunkumar2017mcmgpu,milic2017numa,hu2025hipkittens,chung2026chiplet} reduce remote traffic through page placement and tile mapping, but they do not account for cross-die speed asymmetry or power-capped frequency throttling. \oursys{} closes these gaps with a full-TMEM wide-tile design, fine-grained asynchronous rotation, and speed-proportional dual-die partitioning, outperforming state-of-the-art libraries across dense GEMM and fused Mega-MoE (\cref{sec:eval_gemm,sec:eval_moe}).

\paragraph{Kernel Auto-Tuning and Schedule Search.}
Empirical auto-tuners~\citep{whaley2001atlas,ansel2014opentuner,nugteren2015cltune,vanwerkhoven2019kerneltuner,tillet2019triton,schoonhoven2022going}, tensor compilers~\citep{chen2018tvm,zheng2020ansor,shao2022metaschedule,wu2025mirage}, and black-box optimizers~\citep{bergstra2011tpe,deb2002nsga2,akiba2019optuna} search over predefined kernel configurations by timing candidates on hardware or evaluating cycle-based cost models. However, their search spaces omit fine-grained tile residency order, variable-width load balancing, and dual-die placement, which caps achievable performance even under exhaustive search (\cref{sec:eval_planner}). Furthermore, short on-device timing runs execute at boost clocks and favor schedules that slow down once $P_{\text{cap}}$ engages, whereas steady-state hardware timing requires seconds per candidate. \oursys{} exposes a parameterized design space, the kernel configuration together with the tile schedule, and pairs it with a CPU-side tile scheduler that synthesizes tile schedules via integer programming and a kernel selection step that ranks the candidates with our power-wall cost model, delivering higher speedups without GPU tuning runs.

\section{Conclusion}
\label{sec:conclusion}

At the power wall, a kernel's runtime is governed by its energy consumption. We present \oursys{}, a framework that uses energy--frequency modeling and power-aware speed limits to guide wide-tile asynchronous kernels and power-aware scheduling beyond latency hiding. Across 60 dense-layer shapes, these designs outperform exhaustively selected cuBLAS algorithms by 4.94\% on B200, 6.64\% on H100 NVL, and 2.12\% on RTX PRO 6000 in geomean. The same principles accelerate fused Mega-MoE on eight B200 GPUs by 2.60\%--5.17\% over the upstream kernel with die-local expert weights. We plan to open source \oursys{} to support the research community and encourage power-wall-aware kernel design and scheduling.

\begin{acks}
Generative AI tools were used to assist with writing and figure preparation. The authors reviewed all content and take full responsibility for it. 
\end{acks}

\label{page:references}
\bibliography{refs}

@article{williams2009roofline,
  author  = {Williams, Samuel and Waterman, Andrew and Patterson, David},
  title   = {Roofline: An Insightful Visual Performance Model for Multicore Architectures},
  journal = {Communications of the ACM},
  volume  = {52},
  number  = {4},
  pages   = {65--76},
  year    = {2009}
}

@inproceedings{hong2010ipp,
  author    = {Hong, Sunpyo and Kim, Hyesoon},
  title     = {An Integrated {GPU} Power and Performance Model},
  booktitle = {Proceedings of the 37th Annual International Symposium on Computer Architecture (ISCA)},
  year      = {2010}
}

@inproceedings{leng2013gpuwattch,
  author    = {Leng, Jingwen and Hetherington, Tayler and ElTantawy, Ahmed and Gilani, Syed and Kim, Nam Sung and
               Aamodt, Tor M. and Reddi, Vijay Janapa},
  title     = {{GPUWattch}: Enabling Energy Optimizations in {GPGPUs}},
  booktitle = {Proceedings of the 40th Annual International Symposium on Computer Architecture (ISCA)},
  year      = {2013}
}

@inproceedings{kandiah2021accelwattch,
  author    = {Kandiah, Vijay and Peverelle, Scott and Khairy, Mahmoud and Pan, Junrui and Manjunath, Amogh and
               Rogers, Timothy G. and Aamodt, Tor M. and Hardavellas, Nikos},
  title     = {{AccelWattch}: A Power Modeling Framework for Modern {GPUs}},
  booktitle = {Proceedings of the 54th IEEE/ACM International Symposium on Microarchitecture (MICRO)},
  year      = {2021}
}

@inproceedings{patel2024polca,
  author    = {Patel, Pratyush and Choukse, Esha and Zhang, Chaojie and Goiri, {\'I}{\~n}igo and Warrier, Brijesh and
               Mahalingam, Nithish and Bianchini, Ricardo},
  title     = {Characterizing Power Management Opportunities for {LLMs} in the Cloud},
  booktitle = {Proceedings of the 29th ACM International Conference on Architectural Support for Programming
               Languages and Operating Systems (ASPLOS)},
  pages     = {207--222},
  year      = {2024},
  doi       = {10.1145/3620666.3651329}
}

@inproceedings{patel2024splitwise,
  author    = {Patel, Pratyush and Choukse, Esha and Zhang, Chaojie and Shah, Aashaka and Goiri, {\'I}{\~n}igo and
               Maleki, Saeed and Bianchini, Ricardo},
  title     = {Splitwise: Efficient Generative {LLM} Inference Using Phase Splitting},
  booktitle = {Proceedings of the 51st Annual International Symposium on Computer Architecture (ISCA)},
  year      = {2024}
}

@misc{he2024predictable,
  author       = {He, Horace},
  title        = {Strangely, Matrix Multiplications on {GPUs} Run Faster When Given ``Predictable'' Data!},
  year         = {2024},
  howpublished = {\url{https://www.thonking.ai/p/strangely-matrix-multiplications}}
}

@inproceedings{hong2009analytical,
  author    = {Hong, Sunpyo and Kim, Hyesoon},
  title     = {An Analytical Model for a {GPU} Architecture with Memory-level and Thread-level Parallelism Awareness},
  booktitle = {Proceedings of the 36th Annual International Symposium on Computer Architecture (ISCA)},
  year      = {2009}
}

@inproceedings{huang2014gpumech,
  author    = {Huang, Jen-Cheng and Lee, Joo Hwan and Kim, Hyesoon and Lee, Hsien-Hsin S.},
  title     = {{GPUMech}: {GPU} Performance Modeling Technique Based on Interval Analysis},
  booktitle = {Proceedings of the 47th IEEE/ACM International Symposium on Microarchitecture (MICRO)},
  year      = {2014}
}

@inproceedings{lee2022gcom,
  author    = {Lee, Jounghoo and Ha, Yeonan and Lee, Suhyun and Woo, Jinyoung and Lee, Jinho and Jang, Hanhwi and
               Kim, Youngsok},
  title     = {{GCoM}: A Detailed {GPU} Core Model for Accurate Analytical Modeling of Modern {GPUs}},
  booktitle = {Proceedings of the 49th Annual International Symposium on Computer Architecture (ISCA)},
  pages     = {424--436},
  year      = {2022},
  doi       = {10.1145/3470496.3527384}
}

@inproceedings{khairy2020accelsim,
  author    = {Khairy, Mahmoud and Shen, Zhesheng and Aamodt, Tor M. and Rogers, Timothy G.},
  title     = {{Accel-Sim}: An Extensible Simulation Framework for Validated {GPU} Modeling},
  booktitle = {Proceedings of the 47th Annual International Symposium on Computer Architecture (ISCA)},
  year      = {2020}
}

@inproceedings{gregersen2024input,
  author    = {Gregersen, Theo and Patel, Pratyush and Choukse, Esha},
  title     = {Input-Dependent Power Usage in {GPUs}},
  booktitle = {SC24-W: Workshops of the International Conference for High Performance Computing, Networking, Storage
               and Analysis (Sustainable Supercomputing)},
  year      = {2024}
}

@inproceedings{arunkumar2019gpujoule,
  author    = {Arunkumar, Akhil and Bolotin, Evgeny and Nellans, David and Wu, Carole-Jean},
  title     = {Understanding the Future of Energy Efficiency in Multi-Module {GPUs}},
  booktitle = {IEEE International Symposium on High Performance Computer Architecture (HPCA)},
  year      = {2019}
}

@inproceedings{tran2026wattchmen,
  author    = {Tran, Brandon and Maiterth, Matthias and Shin, Woong and Sinclair, Matthew D. and Venkataraman, Shivaram},
  title     = {Wattchmen: Watching the Wattchers -- High Fidelity, Flexible {GPU} Energy Modeling},
  booktitle = {Proceedings of the 40th ACM International Conference on Supercomputing (ICS)},
  pages     = {145--160},
  year      = {2026},
  doi       = {10.1145/3797905.3800514}
}

@inproceedings{lee2026energaizer,
  author    = {Lee, Kyungmi and Song, Zhiye and Lee, Eun Kyung and Zhang, Xin and Eilam, Tamar and Chandrakasan, Anantha P.},
  title     = {{EnergAIzer}: Fast and Accurate {GPU} Power Estimation Framework for {AI} Workloads},
  booktitle = {IEEE International Symposium on Performance Analysis of Systems and Software (ISPASS)},
  pages     = {435--447},
  year      = {2026},
  doi       = {10.1109/ISPASS69572.2026.00049}
}

@article{guerreiro2019decoupling,
  author  = {Guerreiro, Jo{\~a}o and Ilic, Aleksandar and Roma, Nuno and Tom{\'a}s, Pedro},
  title   = {Modeling and Decoupling the {GPU} Power Consumption for Cross-Domain {DVFS}},
  journal = {IEEE Transactions on Parallel and Distributed Systems},
  volume  = {30},
  number  = {11},
  year    = {2019}
}

@inproceedings{schoonhoven2022going,
  author    = {Schoonhoven, Richard and Veenboer, Bram and van Werkhoven, Ben and Batenburg, K. Joost},
  title     = {Going Green: Optimizing {GPUs} for Energy Efficiency through Model-Steered Auto-Tuning},
  booktitle = {IEEE/ACM International Workshop on Performance Modeling, Benchmarking and Simulation of High
               Performance Computer Systems (PMBS)},
  year      = {2022}
}

@inproceedings{su2014ppep,
  author    = {Su, Bo and Gu, Junli and Shen, Li and Huang, Wei and Greathouse, Joseph L. and Wang, Zhiying},
  title     = {{PPEP}: Online Performance, Power, and Energy Prediction Framework and {DVFS} Space Exploration},
  booktitle = {Proceedings of the 47th IEEE/ACM International Symposium on Microarchitecture (MICRO)},
  pages     = {445--457},
  year      = {2014}
}

@inproceedings{you2023zeus,
  author    = {You, Jie and Chung, Jae-Won and Chowdhury, Mosharaf},
  title     = {Zeus: Understanding and Optimizing {GPU} Energy Consumption of {DNN} Training},
  booktitle = {USENIX Symposium on Networked Systems Design and Implementation (NSDI)},
  year      = {2023}
}

@inproceedings{choi2023envpipe,
  author    = {Choi, Sangjin and Koo, Inhoe and Ahn, Jeongseob and Jeon, Myeongjae and Kwon, Youngjin},
  title     = {{EnvPipe}: Performance-Preserving {DNN} Training Framework for Saving Energy},
  booktitle = {USENIX Annual Technical Conference (ATC)},
  year      = {2023}
}

@article{luo2026dissecting,
  author  = {Luo, Weile and Fan, Ruibo and Li, Zeyu and Du, Dayou and Liu, Hongyuan and Wang, Qiang and Chu, Xiaowen},
  title   = {Dissecting the {NVIDIA} {Hopper} Architecture through Microbenchmarking and Multiple Level Analysis},
  journal = {ACM Transactions on Computer Systems},
  year    = {2026},
  doi     = {10.1145/3856813}
}

@inproceedings{shah2024flashattention3,
  author    = {Shah, Jay and Bikshandi, Ganesh and Zhang, Ying and Thakkar, Vijay and Ramani, Pradeep and Dao, Tri},
  title     = {{FlashAttention-3}: Fast and Accurate Attention with Asynchrony and Low-precision},
  booktitle = {Advances in Neural Information Processing Systems (NeurIPS)},
  year      = {2024}
}

@misc{chang2024flux,
  author       = {Chang, Li-Wen and Bao, Wenlei and Hou, Qi and Jiang, Chengquan and Zheng, Ningxin and Zhong, Yinmin and
                  Zhang, Xuanrun and Song, Zuquan and Yao, Chengji and Jiang, Ziheng and Lin, Haibin and Jin, Xin and Liu, Xin},
  title        = {{FLUX}: Fast Software-based Communication Overlap On {GPUs} Through Kernel Fusion},
  year         = {2024},
  howpublished = {arXiv:2406.06858}
}

@misc{spector2025megakernel,
  author       = {Spector, Benjamin and Juravsky, Jordan and Sul, Stuart and Dugan, Owen and Lim, Dylan and Fu, Dan and
                  Arora, Simran and R{\'e}, Chris},
  title        = {Look Ma, No Bubbles! Designing a Low-Latency Megakernel for {Llama-1B}},
  year         = {2025},
  howpublished = {Hazy Research blog, \url{https://hazyresearch.stanford.edu/blog/2025-05-27-no-bubbles}}
}

@inproceedings{cheng2025mpk,
  author    = {Cheng, Xinhao and Zhang, Zhihao and Zhou, Yu and Ji, Jianan and Jiang, Jinchen and Zhao, Zepeng and Xiao, Ziruo and
               Ye, Zihao and Huang, Yingyi and Lai, Ruihang and Jin, Hongyi and Hou, Bohan and Wu, Mengdi and Dong, Yixin and
               Yip, Anthony and Ye, Zihao and Wang, Songting and Yang, Wenqin and Miao, Xupeng and Chen, Tianqi and Jia, Zhihao},
  title     = {{MPK}: A Compiler and Runtime for Mega-Kernelizing Tensor Programs},
  booktitle = {20th USENIX Symposium on Operating Systems Design and Implementation (OSDI)},
  pages     = {1909--1926},
  year      = {2026},
  note      = {arXiv:2512.22219}
}

@misc{mo2026tilesight,
  author       = {Mo, Zhiwen and Cheng, Yu and Wang, Lei and Tang, Zhengju and Xu, Lei and Li, Guoyu and Dong, Yuqi and
                  Ma, Lingxiao and Xia, Yuqing and Xue, Jilong and Yang, Fan and Mai, Luo and Yang, Zhi and Luk, Wayne and
                  Fan, Hongxiang},
  title        = {{TileSight}: A First-Principles Tile-Centric Analytical {GPU} Performance Model from Cores to Clusters},
  year         = {2026},
  howpublished = {arXiv:2607.22432}
}

@misc{deepgemm,
  author       = {{DeepSeek-AI}},
  title        = {{DeepGEMM}: Clean and Efficient {FP8} {GEMM} Kernels with Fine-Grained Scaling},
  year         = {2025},
  howpublished = {\url{https://github.com/deepseek-ai/DeepGEMM}},
}

@article{chandrakasan1992lowpower,
  author  = {Chandrakasan, Anantha P. and Sheng, Samuel and Brodersen, Robert W.},
  title   = {Low-Power {CMOS} Digital Design},
  journal = {IEEE Journal of Solid-State Circuits},
  volume  = {27},
  number  = {4},
  pages   = {473--484},
  year    = {1992},
  doi     = {10.1109/4.126534}
}

@book{lawson1995least,
  author    = {Lawson, Charles L. and Hanson, Richard J.},
  title     = {Solving Least Squares Problems},
  publisher = {Society for Industrial and Applied Mathematics},
  series    = {Classics in Applied Mathematics},
  volume    = {15},
  year      = {1995},
  doi       = {10.1137/1.9781611971217}
}

@misc{fan2026diescaling,
  title = {Dissecting How Die Scaling Breaks {GPU} Fine-grained Scheduling},
  author = {Fan, Xiaoze and Wang, Jianhao and Cui, Weihao and Zhao, Han and Huang, Zhuobin and Zhou, Yangjie and
            Qiu, Yuxian and Sun, Shixuan and He, Bingsheng and Chen, Quan and Guo, Minyi},
  year = {2026},
  howpublished = {arXiv:2609.24270},
}

@misc{semianalysis2026blackwell,
  title = {Dissecting {Nvidia} {Blackwell} -- Tensor Cores, {PTX} Instructions, {SASS}, Floorsweep, Yield},
  author = {Chen, Kimbo and Patel, Dylan},
  year = {2026},
  howpublished = {SemiAnalysis, \url{https://newsletter.semianalysis.com/p/dissecting-nvidia-blackwell-tensor}},
}

@misc{pan2026async,
  title = {Architecting the Next Generation of Asynchronous, Distributed {GPUs} for the {AI} Era},
  author = {Pan, Junrui and An, Weili and Avalos Baddouh, Cesar and Bose, Christin David and Kang, Ni and
            Barnes, Aaron and Alawneh, Ahmad and Shen, Fangjia and Liu, Yechen and Nallathambi, Anusuya and
            Chandrashekar, Atthin and Rogers, Timothy G.},
  year = {2026},
  howpublished = {arXiv:2608.22602},
}

@inproceedings{bose2026tma,
  title     = {When and Why to Use {TMA}: A Workload-Grounded Characterization of Tensor Memory Acceleration on
               {NVIDIA} {Hopper} {GPUs}},
  author    = {Bose, Christin David and Avalos Baddouh, Cesar and Pan, Junrui and Rogers, Timothy G.},
  booktitle = {IEEE International Symposium on Workload Characterization (IISWC)},
  pages     = {351--362},
  year      = {2026},
  doi       = {10.1109/IISWC71020.2026.00039}
}

@misc{kim2026powerroofline,
  title = {Power Roofline},
  author = {Kim, Michael Jaemin and Yoon, DoeHyun and Stojkovic, Jovan and Zuo, Gefei and Lim, JongBin and
            Jagannati, Harsha and Shu, Zhan and Kanaujia, Shobhit},
  year = {2026},
  howpublished = {\url{https://jovans2.github.io/files/Power_Roofline_MICRO.pdf}},
}

@inproceedings{spector2025thunderkittens,
  title     = {{ThunderKittens}: Simple, Fast, and \emph{Adorable} Kernels},
  author    = {Spector, Benjamin F. and Arora, Simran and Singhal, Aaryan and Parthasarathy, Arjun and Fu, Daniel Y. and
               R{\'e}, Christopher},
  booktitle = {International Conference on Learning Representations (ICLR)},
  pages     = {2097--2130},
  year      = {2025}
}

@misc{nvidia2026locality,
  title = {{CUDA} Programming Guide: Locality Domains},
  author = {{NVIDIA}},
  year = {2026},
  howpublished = {\url{https://docs.nvidia.com/cuda/cuda-programming-guide/04-special-topics/locality-domains.html}},
}

@inproceedings{hong1981io,
  author    = {Hong, Jia-Wei and Kung, H. T.},
  title     = {{I/O} Complexity: The Red-Blue Pebble Game},
  booktitle = {Proceedings of the 13th Annual ACM Symposium on Theory of Computing (STOC)},
  pages     = {326--333},
  year      = {1981}
}

@misc{smith2017tight,
  author       = {Smith, Tyler Michael and Lowery, Bradley and Langou, Julien and van de Geijn, Robert A.},
  title        = {A Tight {I/O} Lower Bound for Matrix Multiplication},
  year         = {2017},
  howpublished = {arXiv:1702.02017}
}

@article{irony2004communication,
  author  = {Irony, Dror and Toledo, Sivan and Tiskin, Alexander},
  title   = {Communication Lower Bounds for Distributed-Memory Matrix Multiplication},
  journal = {Journal of Parallel and Distributed Computing},
  volume  = {64},
  number  = {9},
  pages   = {1017--1026},
  year    = {2004}
}

@misc{hu2025hipkittens,
  author       = {Hu, William and Wadsworth, Drew and Siddens, Sean and Winata, Stanley and Fu, Daniel Y. and
                  Swann, Ryann and Osama, Muhammad and R{\'e}, Christopher and Arora, Simran},
  title        = {{HipKittens}: Fast and Furious {AMD} Kernels},
  year         = {2025},
  howpublished = {arXiv:2511.08083}
}

@misc{chung2026chiplet,
  author       = {Chung, Euijun and Kim, Hyesoon},
  title        = {A Fast Locality Simulator for {GEMM} Design-Space Exploration on Multi-Chiplet {GPUs}},
  year         = {2026},
  howpublished = {arXiv:2606.11716}
}

@misc{kuruppu2026cluster,
  author       = {Kuruppu, Supun and Dangi, Pranav and Li, Wei and Bai, Zhenyu and Mitra, Tulika},
  title        = {Demystifying the Impact of Physical Non-Idealities on Kernel Performance},
  year         = {2026},
  howpublished = {arXiv:2610.37466}
}

@article{theil1961pure,
  author  = {Theil, H. and Goldberger, A. S.},
  title   = {On Pure and Mixed Statistical Estimation in Economics},
  journal = {International Economic Review},
  volume  = {2},
  number  = {1},
  pages   = {65--78},
  year    = {1961}
}

@article{golub1973varpro,
  author  = {Golub, G. H. and Pereyra, V.},
  title   = {The Differentiation of Pseudo-Inverses and Nonlinear Least Squares Problems Whose Variables Separate},
  journal = {SIAM Journal on Numerical Analysis},
  volume  = {10},
  number  = {2},
  pages   = {413--432},
  year    = {1973}
}

@article{stark1995bvls,
  author  = {Stark, Philip B. and Parker, Robert L.},
  title   = {Bounded-Variable Least-Squares: An Algorithm and Applications},
  journal = {Computational Statistics},
  volume  = {10},
  pages   = {129--141},
  year    = {1995}
}

@article{beaumont2002partitioning,
  author  = {Beaumont, Olivier and Boudet, Vincent and Rastello, Fabrice and Robert, Yves},
  title   = {Partitioning a Square into Rectangles: {NP}-Completeness and Approximation Algorithms},
  journal = {Algorithmica},
  volume  = {34},
  number  = {3},
  pages   = {217--239},
  year    = {2002}
}

@inproceedings{aldaas2022tight,
  author    = {Al Daas, Hussam and Ballard, Grey and Grigori, Laura and Kumar, Suraj and Rouse, Kathryn},
  title     = {Brief Announcement: Tight Memory-Independent Parallel Matrix Multiplication Communication Lower Bounds},
  booktitle = {Proceedings of the 34th ACM Symposium on Parallelism in Algorithms and Architectures (SPAA)},
  pages     = {445--448},
  year      = {2022},
  doi       = {10.1145/3490148.3538552}
}

@misc{nvidia2026cublas,
  author       = {{NVIDIA}},
  title        = {{cuBLAS} Library Documentation},
  year         = {2026},
  howpublished = {\url{https://docs.nvidia.com/cuda/cublas/}}
}

@misc{nvidia2026cutlass,
  author       = {{NVIDIA}},
  title        = {{CUTLASS}: {CUDA} Templates and {Python} {DSLs} for High-Performance Linear Algebra},
  year         = {2026},
  howpublished = {\url{https://github.com/NVIDIA/cutlass}}
}

@misc{nvidia2026ncu,
  author       = {{NVIDIA}},
  title        = {{Nsight Compute} Profiling Guide},
  year         = {2026},
  howpublished = {\url{https://docs.nvidia.com/nsight-compute/ProfilingGuide/index.html}}
}

@misc{pytorch2024pingpong,
  author       = {{PyTorch}},
  title        = {Deep Dive on {CUTLASS} {Ping-Pong} {GEMM} Kernel},
  year         = {2024},
  howpublished = {PyTorch Blog, \url{https://pytorch.org/blog/cutlass-ping-pong-gemm-kernel/}}
}

@misc{yang2025qwen3,
  author       = {Yang, An and Li, Anfeng and Yang, Baosong and Zhang, Beichen and Hui, Binyuan and Zheng, Bo and Yu, Bowen and
                  Gao, Chang and Huang, Chengen and Lv, Chenxu and Zheng, Chujie and Liu, Dayiheng and Zhou, Fan and Huang, Fei and
                  Hu, Feng and Ge, Hao and Wei, Haoran and Lin, Huan and Tang, Jialong and Yang, Jian and Tu, Jianhong and
                  Zhang, Jianwei and Yang, Jianxin and Yang, Jiaxi and Zhou, Jing and Zhou, Jingren and Lin, Junyang and Dang, Kai and
                  Bao, Keqin and Yang, Kexin and Yu, Le and Deng, Lianghao and Li, Mei and Xue, Mingfeng and Li, Mingze and
                  Zhang, Pei and Wang, Peng and Zhu, Qin and Men, Rui and Gao, Ruize and Liu, Shixuan and Luo, Shuang and
                  Li, Tianhao and Tang, Tianyi and Yin, Wenbiao and Ren, Xingzhang and Wang, Xinyu and Zhang, Xinyu and
                  Ren, Xuancheng and Fan, Yang and Su, Yang and Zhang, Yichang and Zhang, Yinger and Wan, Yu and Liu, Yuqiong and
                  Wang, Zekun and Cui, Zeyu and Zhang, Zhenru and Zhou, Zhipeng and Qiu, Zihan},
  title        = {{Qwen3} Technical Report},
  year         = {2025},
  howpublished = {arXiv:2505.09388}
}

@misc{alibaba2026qwen38max,
  author       = {{Alibaba Group}},
  title        = {Alibaba Unveils {Qwen3.8-Max}: Its Largest and Most Capable Flagship Model to Date},
  year         = {2026},
  howpublished = {\url{https://www.alibabagroup.com/document-2021044032125272064}}
}

@misc{deepseek2026v4,
  author       = {{DeepSeek-AI}},
  title        = {{DeepSeek-V4}: Towards Highly Efficient Million-Token Context Intelligence},
  year         = {2026},
  howpublished = {arXiv:2606.19348}
}

@misc{rouhani2023microscaling,
  author       = {Darvish Rouhani, Bita and Zhao, Ritchie and More, Ankit and Hall, Mathew and Khodamoradi, Alireza and
                  Deng, Summer and Choudhary, Dhruv and Cornea, Marius and Dellinger, Eric and Denolf, Kristof and
                  Stosic, Dusan and Elango, Venmugil and Golub, Maximilian and Heinecke, Alexander and James-Roxby, Phil and
                  Jani, Dharmesh and Kolhe, Gaurav and Langhammer, Martin and Li, Ada and Melnick, Levi and
                  Mesmakhosroshahi, Maral and Rodriguez, Andres and Schulte, Michael and Shafipour, Rasoul and Shao, Lei and
                  Siu, Michael and Dubey, Pradeep and Micikevicius, Paulius and Naumov, Maxim and Verrilli, Colin and
                  Wittig, Ralph and Burger, Doug and Chung, Eric},
  title        = {Microscaling Data Formats for Deep Learning},
  year         = {2023},
  howpublished = {arXiv:2310.10537}
}

@inproceedings{arunkumar2017mcmgpu,
  author    = {Arunkumar, Akhil and Bolotin, Evgeny and Cho, Benjamin and Milic, Ugljesa and Ebrahimi, Eiman and
               Villa, Oreste and Jaleel, Aamer and Wu, Carole-Jean and Nellans, David},
  title     = {{MCM-GPU}: Multi-Chip-Module {GPUs} for Continued Performance Scalability},
  booktitle = {Proceedings of the 44th Annual International Symposium on Computer Architecture (ISCA)},
  year      = {2017}
}

@misc{lam2025b200,
  author       = {Lam, Chester},
  title        = {{Nvidia's} {B200}: Keeping the {CUDA} Juggernaut Rolling},
  year         = {2025},
  howpublished = {Chips and Cheese, \url{https://chipsandcheese.com/p/nvidias-b200-keeping-the-cuda-juggernaut}}
}

@misc{zadouri2026flashattention4,
  author       = {Zadouri, Ted and Hoehnerbach, Markus and Shah, Jay and Liu, Timmy and Thakkar, Vijay and Dao, Tri},
  title        = {{FlashAttention-4}: Algorithm and Kernel Pipelining Co-Design for Asymmetric Hardware Scaling},
  year         = {2026},
  howpublished = {arXiv:2603.05451}
}

@misc{daoailab2026quack,
  author       = {{Dao-AILab}},
  title        = {{QuACK}: A Quirky Assortment of {CuTe} Kernels},
  year         = {2026},
  howpublished = {\url{https://github.com/Dao-AILab/quack}}
}

@book{brent1973algorithms,
  author    = {Brent, Richard P.},
  title     = {Algorithms for Minimization without Derivatives},
  publisher = {Prentice-Hall},
  address   = {Englewood Cliffs, NJ},
  year      = {1973}
}

@article{huangfu2018highs,
  author  = {Huangfu, Q. and Hall, J. A. J.},
  title   = {Parallelizing the Dual Revised Simplex Method},
  journal = {Mathematical Programming Computation},
  volume  = {10},
  number  = {1},
  pages   = {119--142},
  year    = {2018},
  doi     = {10.1007/s12532-017-0130-5}
}

@article{raffel2020t5,
  author  = {Raffel, Colin and Shazeer, Noam and Roberts, Adam and Lee, Katherine and Narang, Sharan and Matena, Michael and
             Zhou, Yanqi and Li, Wei and Liu, Peter J.},
  title   = {Exploring the Limits of Transfer Learning with a Unified Text-to-Text Transformer},
  journal = {Journal of Machine Learning Research},
  volume  = {21},
  number  = {140},
  pages   = {1--67},
  year    = {2020}
}

@inproceedings{burtscher2014measuring,
  author    = {Burtscher, Martin and Zecena, Ivan and Zong, Ziliang},
  title     = {Measuring {GPU} Power with the {K20} Built-in Sensor},
  booktitle = {Proceedings of the Workshop on General Purpose Processing Using GPUs (GPGPU)},
  pages     = {28--36},
  year      = {2014},
  doi       = {10.1145/2576779.2576783}
}

@inproceedings{yang2024accurate,
  author    = {Yang, Zeyu and Adamek, Karel and Armour, Wesley},
  title     = {Accurate and Convenient Energy Measurements for {GPUs}: A Detailed Study of {NVIDIA} {GPU}'s Built-In Power Sensor},
  booktitle = {Proceedings of the International Conference for High Performance Computing, Networking, Storage and
               Analysis (SC)},
  pages     = {1--17},
  year      = {2024},
  doi       = {10.1109/SC41406.2024.00028}
}

@inproceedings{chung2024perseus,
  author    = {Chung, Jae-Won and Gu, Yile and Jang, Insu and Meng, Luoxi and Bansal, Nikhil and Chowdhury, Mosharaf},
  title     = {Reducing Energy Bloat in Large Model Training},
  booktitle = {Proceedings of the ACM SIGOPS 30th Symposium on Operating Systems Principles (SOSP)},
  pages     = {144--159},
  year      = {2024},
  doi       = {10.1145/3694715.3695970}
}

@inproceedings{stojkovic2025dynamollm,
  author    = {Stojkovic, Jovan and Zhang, Chaojie and Goiri, {\'I}{\~n}igo and Torrellas, Josep and Choukse, Esha},
  title     = {{DynamoLLM}: Designing {LLM} Inference Clusters for Performance and Energy Efficiency},
  booktitle = {IEEE International Symposium on High Performance Computer Architecture (HPCA)},
  pages     = {1348--1362},
  year      = {2025},
  doi       = {10.1109/HPCA61900.2025.00102}
}

@inproceedings{zhao2023sustainable,
  author    = {Zhao, Dan and Samsi, Siddharth and McDonald, Joseph and Li, Baolin and Bestor, David and Jones, Michael and
               Tiwari, Devesh and Gadepally, Vijay},
  title     = {Sustainable Supercomputing for {AI}: {GPU} Power Capping at {HPC} Scale},
  booktitle = {Proceedings of the 2023 ACM Symposium on Cloud Computing (SoCC)},
  pages     = {588--596},
  year      = {2023},
  doi       = {10.1145/3620678.3624793}
}

@inproceedings{choi2013roofline,
  author    = {Choi, Jee Whan and Bedard, Daniel and Fowler, Robert and Vuduc, Richard},
  title     = {A Roofline Model of Energy},
  booktitle = {IEEE International Symposium on Parallel and Distributed Processing (IPDPS)},
  pages     = {661--672},
  year      = {2013},
  doi       = {10.1109/IPDPS.2013.77}
}

@article{ilic2017beyond,
  author  = {Ilic, Aleksandar and Pratas, Frederico and Sousa, Leonel},
  title   = {Beyond the Roofline: Cache-Aware Power and Energy-Efficiency Modeling for Multi-Cores},
  journal = {IEEE Transactions on Computers},
  volume  = {66},
  number  = {1},
  pages   = {52--58},
  year    = {2017}
}

@inproceedings{bakhoda2009gpgpusim,
  author    = {Bakhoda, Ali and Yuan, George L. and Fung, Wilson W. L. and Wong, Henry and Aamodt, Tor M.},
  title     = {Analyzing {CUDA} Workloads Using a Detailed {GPU} Simulator},
  booktitle = {IEEE International Symposium on Performance Analysis of Systems and Software (ISPASS)},
  pages     = {163--174},
  year      = {2009},
  doi       = {10.1109/ISPASS.2009.4919648}
}

@inproceedings{sun2019mgpusim,
  author    = {Sun, Yifan and Baruah, Trinayan and Mojumder, Saiful A. and Dong, Shi and Gong, Xiang and Treadway, Shane and
               Bao, Yuhui and Hance, Spencer and McCardwell, Carter and Zhao, Vincent and Barclay, Harrison and
               Ziabari, Amir Kavyan and Chen, Zhongliang and Ubal, Rafael and Abell{\'a}n, Jos{\'e} L. and Kim, John and
               Joshi, Ajay and Kaeli, David},
  title     = {{MGPUSim}: Enabling Multi-{GPU} Performance Modeling and Optimization},
  booktitle = {Proceedings of the 46th International Symposium on Computer Architecture (ISCA)},
  pages     = {197--209},
  year      = {2019},
  doi       = {10.1145/3307650.3322230}
}

@inproceedings{lee2025neusight,
  author    = {Lee, Seonho and Phanishayee, Amar and Mahajan, Divya},
  title     = {Forecasting {GPU} Performance for Deep Learning Training and Inference},
  booktitle = {Proceedings of the 30th ACM International Conference on Architectural Support for Programming Languages
               and Operating Systems (ASPLOS)},
  pages     = {493--508},
  year      = {2025},
  doi       = {10.1145/3669940.3707265}
}

@article{ballard2011minimizing,
  author  = {Ballard, Grey and Demmel, James and Holtz, Olga and Schwartz, Oded},
  title   = {Minimizing Communication in Numerical Linear Algebra},
  journal = {SIAM Journal on Matrix Analysis and Applications},
  volume  = {32},
  number  = {3},
  pages   = {866--901},
  year    = {2011},
  doi     = {10.1137/090769156}
}

@inproceedings{olivry2020automated,
  author    = {Olivry, Auguste and Langou, Julien and Pouchet, Louis-No{\"e}l and Sadayappan, P. and Rastello, Fabrice},
  title     = {Automated Derivation of Parametric Data Movement Lower Bounds for Affine Programs},
  booktitle = {Proceedings of the 41st ACM SIGPLAN Conference on Programming Language Design and Implementation (PLDI)},
  pages     = {808--822},
  year      = {2020},
  doi       = {10.1145/3385412.3385989}
}

@article{goto2008anatomy,
  author  = {Goto, Kazushige and van de Geijn, Robert A.},
  title   = {Anatomy of High-Performance Matrix Multiplication},
  journal = {ACM Transactions on Mathematical Software},
  volume  = {34},
  number  = {3},
  pages   = {1--25},
  year    = {2008},
  doi     = {10.1145/1356052.1356053}
}

@article{vanzee2015blis,
  author  = {Van Zee, Field G. and van de Geijn, Robert A.},
  title   = {{BLIS}: A Framework for Rapidly Instantiating {BLAS} Functionality},
  journal = {ACM Transactions on Mathematical Software},
  volume  = {41},
  number  = {3},
  pages   = {1--33},
  year    = {2015},
  doi     = {10.1145/2764454}
}

@article{tomov2010magma,
  author  = {Tomov, Stanimire and Dongarra, Jack and Baboulin, Marc},
  title   = {Towards Dense Linear Algebra for Hybrid {GPU} Accelerated Manycore Systems},
  journal = {Parallel Computing},
  volume  = {36},
  number  = {5--6},
  pages   = {232--240},
  year    = {2010},
  doi     = {10.1016/j.parco.2009.12.005}
}

@misc{micikevicius2022fp8,
  author       = {Micikevicius, Paulius and Stosic, Dusan and Burgess, Neil and Cornea, Marius and Dubey, Pradeep and
                  Grisenthwaite, Richard and Ha, Sangwon and Heinecke, Alexander and Judd, Patrick and Kamalu, John and
                  Mellempudi, Naveen and Oberman, Stuart and Shoeybi, Mohammad and Siu, Michael and Wu, Hao},
  title        = {{FP8} Formats for Deep Learning},
  year         = {2022},
  howpublished = {arXiv:2209.05433}
}

@article{whaley2001atlas,
  author  = {Whaley, R. Clint and Petitet, Antoine and Dongarra, Jack J.},
  title   = {Automated Empirical Optimizations of Software and the {ATLAS} Project},
  journal = {Parallel Computing},
  volume  = {27},
  number  = {1--2},
  pages   = {3--35},
  year    = {2001},
  doi     = {10.1016/S0167-8191(00)00087-9}
}

@article{frigo2005fftw3,
  author  = {Frigo, Matteo and Johnson, Steven G.},
  title   = {The Design and Implementation of {FFTW3}},
  journal = {Proceedings of the IEEE},
  volume  = {93},
  number  = {2},
  pages   = {216--231},
  year    = {2005},
  doi     = {10.1109/JPROC.2004.840301}
}

@inproceedings{ansel2014opentuner,
  author    = {Ansel, Jason and Kamil, Shoaib and Veeramachaneni, Kalyan and Ragan-Kelley, Jonathan and Bosboom, Jeffrey and
               O'Reilly, Una-May and Amarasinghe, Saman},
  title     = {{OpenTuner}: An Extensible Framework for Program Autotuning},
  booktitle = {Proceedings of the 23rd International Conference on Parallel Architectures and Compilation (PACT)},
  pages     = {303--316},
  year      = {2014},
  doi       = {10.1145/2628071.2628092}
}

@article{vanwerkhoven2019kerneltuner,
  author  = {van Werkhoven, Ben},
  title   = {Kernel Tuner: A Search-Optimizing {GPU} Code Auto-Tuner},
  journal = {Future Generation Computer Systems},
  volume  = {90},
  pages   = {347--358},
  year    = {2019},
  doi     = {10.1016/j.future.2018.08.004}
}

@inproceedings{nugteren2015cltune,
  author    = {Nugteren, Cedric and Codreanu, Valeriu},
  title     = {{CLTune}: A Generic Auto-Tuner for {OpenCL} Kernels},
  booktitle = {IEEE 9th International Symposium on Embedded Multicore/Many-core Systems-on-Chip (MCSoC)},
  pages     = {195--202},
  year      = {2015},
  doi       = {10.1109/MCSoC.2015.10}
}

@inproceedings{chen2018tvm,
  author    = {Chen, Tianqi and Moreau, Thierry and Jiang, Ziheng and Zheng, Lianmin and Yan, Eddie and Shen, Haichen and
               Cowan, Meghan and Wang, Leyuan and Hu, Yuwei and Ceze, Luis and Guestrin, Carlos and Krishnamurthy, Arvind},
  title     = {{TVM}: An Automated End-to-End Optimizing Compiler for Deep Learning},
  booktitle = {13th USENIX Symposium on Operating Systems Design and Implementation (OSDI)},
  pages     = {578--594},
  year      = {2018}
}

@inproceedings{chen2018learning,
  author    = {Chen, Tianqi and Zheng, Lianmin and Yan, Eddie and Jiang, Ziheng and Moreau, Thierry and Ceze, Luis and
               Guestrin, Carlos and Krishnamurthy, Arvind},
  title     = {Learning to Optimize Tensor Programs},
  booktitle = {Advances in Neural Information Processing Systems (NeurIPS)},
  volume    = {31},
  year      = {2018}
}

@inproceedings{zheng2020ansor,
  author    = {Zheng, Lianmin and Jia, Chengfan and Sun, Minmin and Wu, Zhao and Yu, Cody Hao and Haj-Ali, Ameer and
               Wang, Yida and Yang, Jun and Zhuo, Danyang and Sen, Koushik and Gonzalez, Joseph E. and Stoica, Ion},
  title     = {Ansor: Generating High-Performance Tensor Programs for Deep Learning},
  booktitle = {14th USENIX Symposium on Operating Systems Design and Implementation (OSDI)},
  pages     = {863--879},
  year      = {2020}
}

@inproceedings{shao2022metaschedule,
  author    = {Shao, Junru and Zhou, Xiyou and Feng, Siyuan and Hou, Bohan and Lai, Ruihang and Jin, Hongyi and Lin, Wuwei and
               Masuda, Masahiro and Yu, Cody Hao and Chen, Tianqi},
  title     = {Tensor Program Optimization with Probabilistic Programs},
  booktitle = {Advances in Neural Information Processing Systems (NeurIPS)},
  volume    = {35},
  pages     = {35783--35796},
  year      = {2022}
}

@inproceedings{tillet2019triton,
  author    = {Tillet, Philippe and Kung, H. T. and Cox, David},
  title     = {Triton: An Intermediate Language and Compiler for Tiled Neural Network Computations},
  booktitle = {Proceedings of the 3rd ACM SIGPLAN International Workshop on Machine Learning and Programming Languages
               (MAPL)},
  pages     = {10--19},
  year      = {2019},
  doi       = {10.1145/3315508.3329973}
}

@inproceedings{wu2025mirage,
  author    = {Wu, Mengdi and Cheng, Xinhao and Liu, Shengyu and Shi, Chunan and Ji, Jianan and Ao, Man Kit and
               Velliengiri, Praveen and Miao, Xupeng and Padon, Oded and Jia, Zhihao},
  title     = {Mirage: A Multi-Level Superoptimizer for Tensor Programs},
  booktitle = {19th USENIX Symposium on Operating Systems Design and Implementation (OSDI)},
  pages     = {21--38},
  year      = {2025}
}

@inproceedings{akiba2019optuna,
  author    = {Akiba, Takuya and Sano, Shotaro and Yanase, Toshihiko and Ohta, Takeru and Koyama, Masanori},
  title     = {Optuna: A Next-Generation Hyperparameter Optimization Framework},
  booktitle = {Proceedings of the 25th ACM SIGKDD International Conference on Knowledge Discovery \& Data Mining (KDD)},
  pages     = {2623--2631},
  year      = {2019},
  doi       = {10.1145/3292500.3330701}
}

@inproceedings{bergstra2011tpe,
  author    = {Bergstra, James and Bardenet, R{\'e}mi and Bengio, Yoshua and K{\'e}gl, Bal{\'a}zs},
  title     = {Algorithms for Hyper-Parameter Optimization},
  booktitle = {Advances in Neural Information Processing Systems (NeurIPS)},
  volume    = {24},
  year      = {2011}
}

@article{deb2002nsga2,
  author  = {Deb, Kalyanmoy and Pratap, Amrit and Agarwal, Sameer and Meyarivan, T.},
  title   = {A Fast and Elitist Multiobjective Genetic Algorithm: {NSGA-II}},
  journal = {IEEE Transactions on Evolutionary Computation},
  volume  = {6},
  number  = {2},
  pages   = {182--197},
  year    = {2002},
  doi     = {10.1109/4235.996017}
}

@inproceedings{milic2017numa,
  author    = {Milic, Ugljesa and Villa, Oreste and Bolotin, Evgeny and Arunkumar, Akhil and Ebrahimi, Eiman and
               Jaleel, Aamer and Ramirez, Alex and Nellans, David},
  title     = {Beyond the Socket: {NUMA}-Aware {GPUs}},
  booktitle = {Proceedings of the 50th Annual IEEE/ACM International Symposium on Microarchitecture (MICRO)},
  pages     = {123--135},
  year      = {2017},
  doi       = {10.1145/3123939.3124534}
}

@inproceedings{dao2022flashattention,
  author    = {Dao, Tri and Fu, Daniel Y. and Ermon, Stefano and Rudra, Atri and R{\'e}, Christopher},
  title     = {{FlashAttention}: Fast and Memory-Efficient Exact Attention with {IO}-Awareness},
  booktitle = {Advances in Neural Information Processing Systems (NeurIPS)},
  volume    = {35},
  pages     = {16344--16359},
  year      = {2022}
}

@inproceedings{gale2023megablocks,
  author    = {Gale, Trevor and Narayanan, Deepak and Young, Cliff and Zaharia, Matei},
  title     = {{MegaBlocks}: Efficient Sparse Training with Mixture-of-Experts},
  booktitle = {Proceedings of Machine Learning and Systems (MLSys)},
  volume    = {5},
  pages     = {288--304},
  year      = {2023}
}

@inproceedings{narayanan2021megatron,
  author    = {Narayanan, Deepak and Shoeybi, Mohammad and Casper, Jared and LeGresley, Patrick and Patwary, Mostofa and
               Korthikanti, Vijay and Vainbrand, Dmitri and Kashinkunti, Prethvi and Bernauer, Julie and Catanzaro, Bryan and
               Phanishayee, Amar and Zaharia, Matei},
  title     = {Efficient Large-Scale Language Model Training on {GPU} Clusters Using {Megatron-LM}},
  booktitle = {Proceedings of the International Conference for High Performance Computing, Networking, Storage and
               Analysis (SC)},
  pages     = {1--15},
  year      = {2021},
  doi       = {10.1145/3458817.3476209}
}

@article{hoerl1970ridge,
  author  = {Hoerl, Arthur E. and Kennard, Robert W.},
  title   = {Ridge Regression: Biased Estimation for Nonorthogonal Problems},
  journal = {Technometrics},
  volume  = {12},
  number  = {1},
  pages   = {55--67},
  year    = {1970},
  doi     = {10.1080/00401706.1970.10488634}
}
\bibliographystyle{ACM-Reference-Format}

\clearpage

\end{document}